\documentclass{article}

\usepackage{microtype}
\usepackage{graphicx}
\usepackage{subcaption}
\usepackage{booktabs} % for professional tables

\usepackage{multirow}

\usepackage{hyperref}

\usepackage[preprint]{icml2026}
\usepackage{amsmath}
\usepackage{amssymb}
\usepackage{mathtools}
\usepackage{amsthm}

\usepackage[capitalize,noabbrev]{cleveref}

\theoremstyle{plain}

\theoremstyle{definition}

\theoremstyle{remark}

\usepackage{enumitem}

\usepackage[textsize=tiny]{todonotes}

\icmltitlerunning{Submission and Formatting Instructions for ICML 2026}

\begin{document}

\twocolumn[
  \icmltitle{Dual-View Predictive Diffusion: Lightweight Speech Enhancement via Spectrogram-Image Synergy}



 %  \icmlsetsymbol{equal}{*}

  \begin{icmlauthorlist}
    \icmlauthor{Ke Xue}{yyy}
    \icmlauthor{Rongfei Fan}{yyy}
    \icmlauthor{Kai Li}{comp}
    \icmlauthor{Shanping Yu}{yyy}
    \icmlauthor{Puning Zhao}{sch}
    \icmlauthor{Jianping An}{yyy}
  \end{icmlauthorlist}

  \icmlaffiliation{yyy}{School of Cyberspace Science and Technology, Beijing Institute of Technology, Beijing 100081, China}
  \icmlaffiliation{comp}{Department of Computer Science and Technology, Institute for AI, BNRist, Tsinghua University, Beijing 100084, China}
  \icmlaffiliation{sch}{School of Cyberspace Science and Technology
Sun Yat-sen University, Guangzhou 510006, China}

  \icmlcorrespondingauthor{Rongfei Fan}{fanrongfei@bit.edu.cn}

  \icmlkeywords{Machine Learning, ICML}

  \vskip 0.3in
]



\printAffiliationsAndNotice{}  % no special notice (required even if empty)

\begin{abstract}
Diffusion models have recently set new benchmarks in Speech Enhancement (SE). However, most existing score-based models treat speech spectrograms merely as generic 2D images, applying uniform processing that ignores the intrinsic structural sparsity of audio, which results in inefficient spectral representation and prohibitive computational complexity. To bridge this gap, we propose \textbf{DVPD}, an extremely lightweight \textbf{D}ual-\textbf{V}iew \textbf{P}redictive \textbf{D}iffusion model, which uniquely exploits the dual nature of spectrograms as both visual textures and physical frequency-domain representations across both training and inference stages. Specifically, during training, we optimize spectral utilization via the Frequency-Adaptive Non-uniform Compression (FANC) encoder, which preserves critical low-frequency harmonics while pruning high-frequency redundancies. Simultaneously, we introduce a Lightweight Image-based Spectro-Awareness (LISA) module to capture features from a visual perspective with minimal overhead. During inference, we propose a Training-free Lossless Boost (TLB) strategy that leverages the same dual-view priors to refine generation quality without any additional fine-tuning. Extensive experiments across various benchmarks demonstrate that DVPD achieves state-of-the-art performance while requiring only \textbf{35\%} of the parameters and \textbf{40\%} of the inference MACs compared to SOTA lightweight model, PGUSE. These results highlight DVPD's superior ability to balance high-fidelity speech quality with extreme architectural efficiency. Code and audio samples are available at the anonymous website: \url{https://anonymous.4open.science/r/dvpd_demo-E630}

\end{abstract}

\section{Introduction}

%In real-world acoustic environments, speech signals are ubiquitously corrupted by noise \cite{braun2021towards}, room reverberation, signal clipping \cite{mack2019declipping}, and various other types of distortions. These degradations severely impair speech quality and intelligibility, hindering effective human-to-human and human-computer communication. Speech Enhancement (SE) aims to mitigate these deleterious effects and reconstruct high-fidelity audio from corrupted signals. Previous SE research \cite{braun2021towards, mack2019declipping, purushothaman2023speech, wang2021towards} has largely focused on task-oriented scenarios, providing specialized solutions for isolated distortions. However, in practical applications, multiple distortions often coexist and interact in complex ways. While Universal Speech Enhancement (USE) \cite{pascual2019towards, nair2021cascaded, serra2022universal, scheibler2024universal}, with ability to handle diverse degradations simultaneously, represents the current frontier of the field, we argue that a truly robust and superior model should be versatile enough to excel across both specialized tasks and universal restoration scenarios. Bridging the gap between task-specific precision and universal robustness is thus a critical objective for next-generation SE systems.

In real-world acoustic environments, speech signals are commonly degraded by noise \cite{braun2021towards}, reverberation, clipping \cite{mack2019declipping}, and other distortions, which severely impair speech quality and intelligibility. Speech enhancement (SE) seeks to recover high-fidelity speech from corrupted signals. Existing SE methods \cite{braun2021towards, mack2019declipping, purushothaman2023speech, wang2021towards} have predominantly focused on task-oriented settings, offering tailored solutions for individual distortions, whereas practical scenarios often involve multiple interacting degradations. Universal speech enhancement (USE) \cite{pascual2019towards, nair2021cascaded, serra2022universal, scheibler2024universal} has emerged as a promising paradigm for addressing diverse distortions within a unified framework; however, reconciling task-specific precision with universal robustness remains a formidable challenge. 
\vspace{-0.2cm}
\begin{figure}[htb]
\begin{minipage}[b]{1.0 \linewidth}
 \centering
 % \hspace{3cm}  % 向you移动7厘米
 \centerline{\includegraphics[width= \columnwidth]{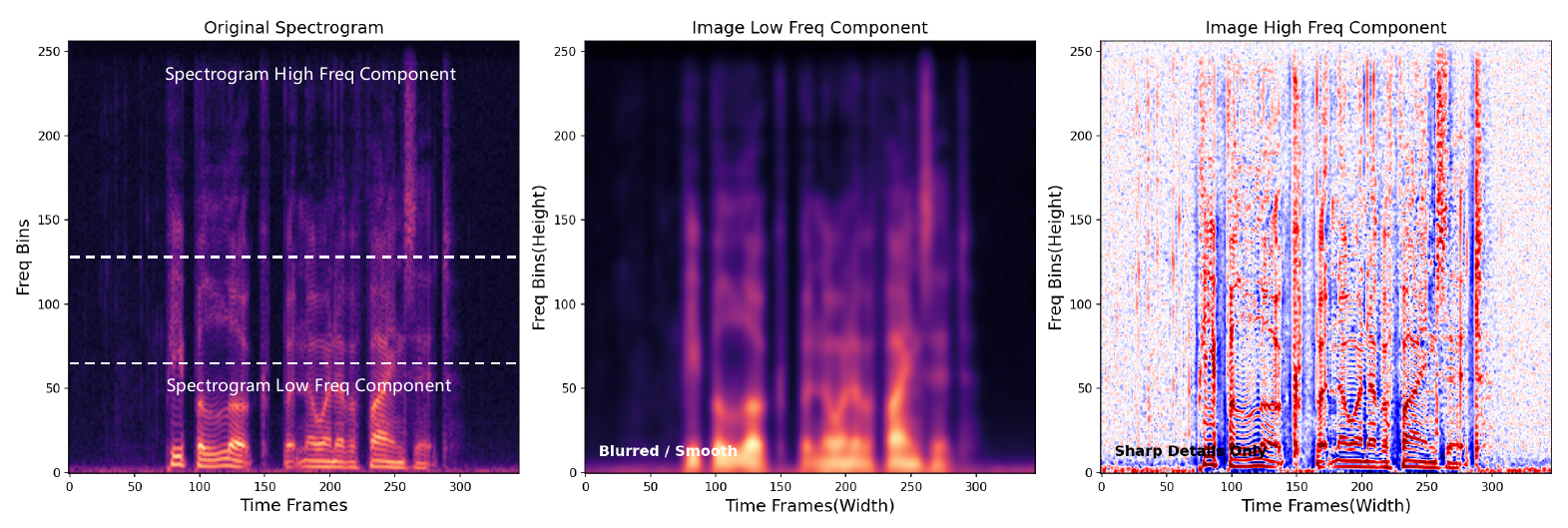}}
\end{minipage}

% \vspace{-0.5cm}
\caption{{Dual-view of spectrogram frequency bands.} Left: Acoustic perspective; Middle and Right: Visual image perspectives of low-frequency and high-frequency components, respectively.}

\label{f:fig1}

\end{figure}
\vspace{-0.2cm}

%Regardless of whether SE or USE is considered, their evolution has been driven by two primary paradigms: predictive (discriminative) and generative methods. Predictive models \cite{krawczyk2014stft, williamson2015complex, park2022manner, lu2023mp, abdulatif2024cmgan} typically treat SE as a regression task, learning a mapping from noisy to clean signals via a ``black-box" optimization of objectives. While these methods are computationally efficient and have achieved significant success in single-distortion tasks, they often suffer from the over-smoothing effect due to the ``mean regression" nature. Consequently, they frequently fail to recover fine-grained acoustic details, resulting in ``muffled" audio and poor generalization in complex, multi-distortion environments.

Current SE methods are primarily bifurcated into two paradigms: predictive and generative. Predictive models \cite{krawczyk2014stft, williamson2015complex, park2022manner, lu2023mp, abdulatif2024cmgan} typically formulate SE as a supervised learning task, optimizing either a direct mapping or a temporal-frequency mask to recover clean speech through deterministic objective functions. Although computational efficient and effective for single distortions, these methods often suffer from over-smoothing due to intrinsic  ``regression to the mean'' phenomenon, leading to the loss of fine-grained acoustic details, “muffled” outputs, and poor generalization in complex multi-distortion scenarios.

In contrast, generative methods model the intrinsic data distribution in a unified latent space, enabling effective recovery under severe information loss. Among various paradigms, including VAEs \cite{kingma2013auto, richter2020speech}, GANs \cite{goodfellow2014generative, kong2020hifi}, and Normalizing Flows \cite{rezende2015variational, yang2025investigating, lee2025flowse}, Diffusion Models \cite{lu2022conditional} have recently emerged as SOTA, driven by their success in image synthesis and their growing adoption in SE.
Score-based diffusion models \cite{welker2022speech, richter2023speech} are particularly notable for formulating diffusion via stochastic differential equations, enabling a continuous-time modeling framework. However, ``pure'' diffusion models entail prohibitive computational costs due to numerous iterative reverse steps.
To mitigate this, recent work has shifted toward a predictive-diffusion paradigm \cite{lemercier2023storm, scheibler2024universal, zhang2025composite}, which employs predictive models as robust priors to constrain the diffusion process, thereby enhancing reconstruction stability while reducing sampling overhead.

Despite these advancements, current methods still remain computationally burdensome. We attribute this inefficiency to a fundamental oversight: spectrograms are treated as generic 2D images with spatially uniform operations, ignoring their underlying physical structure. As illustrated in Fig.~\ref{f:fig1}, we argue that extreme efficiency requires embracing the dual nature of spectrograms from two complementary perspectives: 1) a visual perspective, where spectrograms exhibit image-like textures and spatial patterns \cite{welker2022speech}; 2) an acoustic perspective, where spectrograms are physical representations with strong anisotropy and non-uniform information density \cite{yu2023efficient}, featuring information-dense low-frequency harmonics and sparse yet critical high-frequency transients.

%Although the predictive-generative paradigm has proven effective, current SOTA methods remain computationally prohibitive for edge-device deployment. We argue that this inefficiency stems from a fundamental oversight: existing models treat spectrograms merely as generic 2D images, applying spatially-uniform operations that fail to exploit the inherent physical structure of speech. As illustrated in Figure~\ref{f:fig1}, we posit that the key to extreme efficiency lies in acknowledging the dual-nature of spectrograms through two synergistic perspectives: 1) The Visual Perspective (Spectrogram as Image) \cite{welker2022speech}: The spectrogram exhibits spatial textures, multi-scale structural patterns, and local continuities. 2) The Acoustic Perspective (Spectrogram as Physical Representation): Unlike natural images, spectrograms possess high structural anisotropy and non-uniform information density \cite{yu2023efficient}. Low-frequency regions are information-dense, characterized by stable harmonic stacks and pitch contours; conversely, high-frequency regions are often sparse but contain critical transient details. 

Based on above complementary views, we introduce DVPD, the dual-view predictive diffusion framework that harmonizes visual structural textures with acoustic physics. To optimize spectral efficiency, we design a frequency-adaptive non-uniform compression (FANC) encoder, which employs heterogeneous kernels to preserve low-frequency harmonic integrity while pruning high-frequency redundancies, in alignment with human auditory frequency resolution.
Furthermore, we incorporate a backbone augmented with a multi-range lightweight image-based spectro-awareness (LISA) module to efficiently capture anisotropic features, including horizontal harmonic correlations and vertical transients, with minimal overhead.
At inference time, we further propose a training-free lossless boost (TLB) strategy that exploits the dual-view structure to recalibrate feature scales, yielding consistent quality improvements without additional training. Extensive experiments demonstrate that DVPD achieves a new efficiency and quality trade-off in generative SE.
Notably, our model attains SOTA performance while using only 35\% of the parameters and 40\% of the MACs of the SOTA lightweight model, PGUSE, thereby establishing a new baseline for efficiency and quality in generative SE. % Audio samples are available at the anonymous website: \url{https://anonymous.4open.science/r/dvpd_demo-E630} % \url{https://dvpd-demo.vercel.app}.
% Our contributions are summarized as follows:
% % \begin{itemize}
% \begin{enumerate}[label=\arabic*., nosep, leftmargin=*]
%     \item We introduce a novel dual-view perspective for spectrogram modeling that bridges acoustic physical priors with visual structural textures, providing a theoretical basis for eliminating architectural redundancy in diffusion models.
    
%     \item Built upon this dual-view insight, we propose {DVPD}, an lightweight predictive-diffusion framework. Through the design of the FANC Encoder and {LISA module}, DVPD achieves SOTA performance with only {35\% of parameters} and {40\% of the MACs} of leading models.
    
%     \item Leveraging the structural properties of our dual-view framework, we further design the TLB strategy during inference, which provides ``lossless'' quality improvements by recalibrating feature maps during the reverse process, requiring zero additional training or fine-tuning.
    
% \item We demonstrate through extensive evaluations that our model excels in both USE and task-oriented SE, outperforming SOTA models in both efficiency and robustness.
% \end{enumerate}
% % \end{itemize}
%  % plug-and-play

\begin{figure*}[t]
\begin{minipage}[b]{1.0 \linewidth}
 \centering
 % \hspace{3cm}  % 向you移动7厘米
 \centerline{\includegraphics[width= \columnwidth]{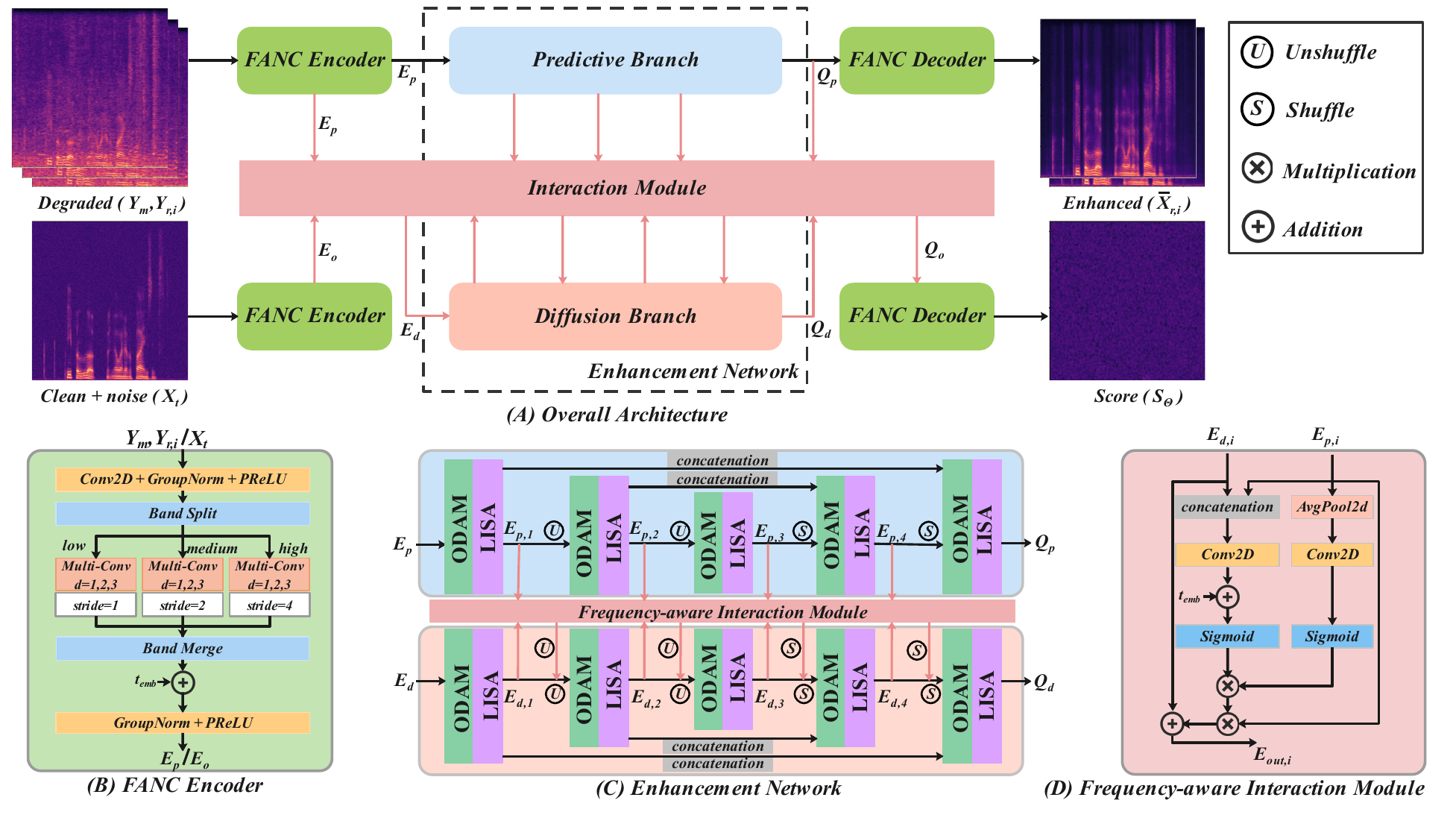}}
\end{minipage}

% \vspace{-0.5cm}
\caption{{Architectural overview of the proposed DVPD.} (A) The overall symmetrical dual-branch structure of DVPD, comprising the predictive branch (upper) and the diffusion branch (lower). (B) Detailed design of the Frequency-Adaptive Non-uniform Compression (FANC) Encoder. (C) Detailed design of enhancement network. (D) Illustration of the Frequency-Aware Interaction module.}

\label{f:fig2}

\end{figure*}

\section{Related Work}

\subsection{Score-based Diffusion Models for SE}
Diffusion Probabilistic Models (DPMs) have revolutionized SE by mitigating the over-smoothing and loss of high-frequency details common in traditional methods \cite{lu2021study, lu2022conditional}. The field has matured with the introduction of continuous-time Stochastic Differential Equations (SDEs), enabling sophisticated score-based modeling in the complex STFT domain \cite{welker2022speech, richter2023speech} and refined conditioning mechanisms \cite{tai2023dose, yang2025dose+}. Despite their generative prowess, these models incur prohibitive computational costs due to the extensive iterative sampling steps. While recent efforts like Latent Diffusion Models (LDM) \cite{zhao2025conditional} attempt to mitigate this by operating in compressed spaces or optimizing trajectories, the inherent inference complexity remains a significant bottleneck.

\subsection{Predictive-Diffusion Hybrid Architectures}
To alleviate the sampling bottleneck and suppress hallucinations, recent trends have converged toward hybrid paradigms that integrate predictive and generative models. These approaches generally follow three trajectories: 1) Conditioning-based integration, where a predictive network serves as an auxiliary conditioner to guide the diffusion process \cite{serra2022universal, scheibler2024universal, kim2024guided}; 2) Cascaded regression refinement, where a predictive model provides an initial estimate or residual, which is then refined by the diffusion branch to reduce the required sampling steps \cite{qiu2023srtnet, lemercier2023storm}; 3) Parallel Architecture, which introduces a dual-branch parallel framework that achieves SOTA performance with relatively low complexity \cite{zhang2025composite}.
DVPD belongs to the parallel category but distinguishes itself through a domain-specific architecture. Instead of relying on generic spatially-homogeneous backbones, we explicitly model ``spectrogram-image synergy'' by bridging acoustic physics with visual structural priors, which allows DVPD to achieve superior performance with significantly fewer parameters and MACs.

\section{The Proposed Model}
\subsection{Preliminaries: Score-based Diffusion Models}

Score-based diffusion models conceptualize SE as a continuous time transformation between clean speech $\mathbf{x}_0 \in \mathbb{R}^{1 \times L}$ and a noisy condition $\mathbf{y} \in \mathbb{R}^{1 \times L}$ via Stochastic Differential Equations (SDEs) \cite{song2020score}. The forward process $\{ \mathbf{x}_t \}_{t=0}^T$ perturbs the data distribution toward a prior distribution, while the reverse process recovers clean signals by learning the score function $\nabla_{\mathbf{x}_t} \log p_t(\mathbf{x}_t)$. In the context of SE, two prominent SDE formulations are widely used: \textit{Ornstein-Uhlenbeck Variance Exploding} (OUVE) \cite{ welker2022speech, richter2023speech} and \textit{Brownian Bridge with Exponential Diffusion} (BBED) \cite{lay2023reducing}.
To mitigate the \textit{prior mismatch} and inference bias inherent in finite-time OUVE formulations, we adopt the BBED. 
Unlike OUVE, where the mean only converges to $\mathbf{y}$ as $T \to \infty$, BBED's linear mean evolution ensures the forward process reaches $\mathbf{y}$ at $T = 1$, significantly enhancing restoration stability. A comprehensive mathematical comparison is provided in \textbf{Appendix \ref{app:diffusion_details}}.

\subsection{Overall Pipeline}

%DVPD employs a parallel predictive-diffusion paradigm, as shown in Figure~\ref{f:fig2}(A), a structural choice inspired by its proven efficacy in balancing deterministic restoration and stochastic refinement \cite{zhang2025composite}. All operations are conducted in the Time-Frequency (T-F) domain using the Short-Time Fourier Transform (STFT).
%Both the predictive and diffusion branches utilize a \textbf{U-Net} foundational backbone. This choice is motivated by U-Net's inherent architectural advantages in spatial feature extraction for image denoising tasks \cite{williams2023unified, si2024freeu, tian2024u}. Furthermore, our proposed {TLB} strategy (see Sec. \ref{sec:inference_strategy}) is specifically engineered on U-Net, making this backbone a critical component of our framework.

DVPD follows the parallel predictive-diffusion architecture illustrated in Fig.~\ref{f:fig2}(A). 
All operations are conducted in the T-F domain using STFT. 
Given a clean waveform $\mathbf{x}_0 \in \mathbb{R}^{1 \times L}$ and a degraded waveform $\mathbf{y} \in \mathbb{R}^{1 \times L}$, we obtain their complex spectrograms 
$\mathbf{X}_{r,i}, \mathbf{Y}_{r,i} \in \mathbb{R}^{2 \times F \times T}$, and the magnitude spectrum of them 
$\mathbf{Y}_m, \mathbf{X}_m \in \mathbb{R}^{1 \times F \times T}$. 
Power-law compression is then applied to compensate for the heavy-tailed distribution of speech amplitudes \cite{gerkmann2010empirical} and improve numerical stability.

\begin{figure*}[t]
\begin{minipage}[b]{1.0 \linewidth}
 \centering
 % \hspace{3cm}  % 向you移动7厘米
 \centerline{\includegraphics[width= \columnwidth]{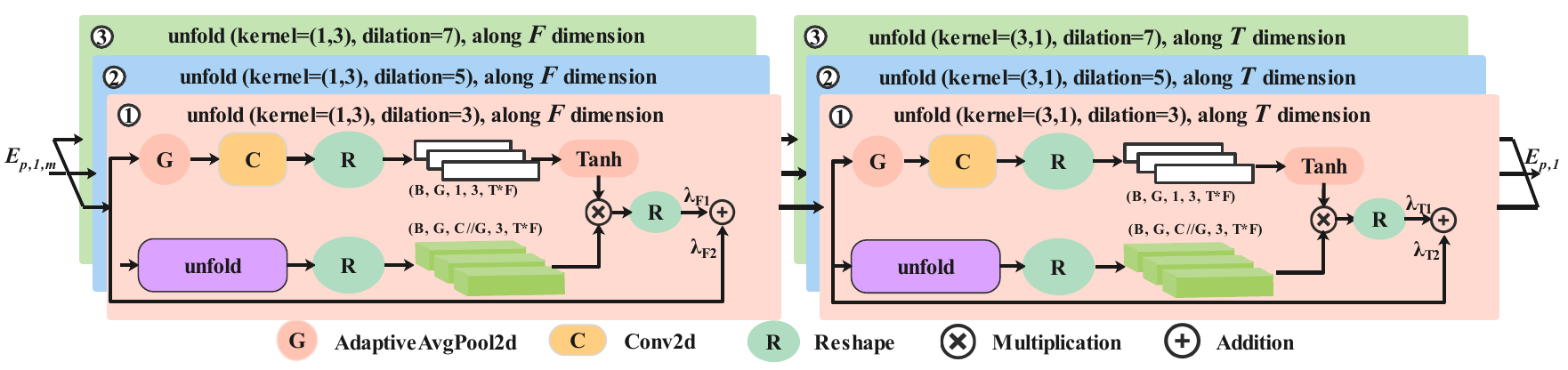}}
\end{minipage}

% \vspace{-0.5cm}
\caption{Detailed architecture of the LISA module.}

\label{f:fig3}

\end{figure*}

The predictive branch takes the degraded spectrogram $\mathbf{Y} = [\mathbf{Y}_{r,i}, \mathbf{Y}_m] \in \mathbb{R}^{3 \times F \times T}$ as input, which is initially projected by the \textit{FANC encoder} into a compressed latent space to yield encoded features $\mathbf{E}_p \in \mathbb{R}^{C \times F_1 \times T}$. Subsequently, $\mathbf{E}_p$ is processed by the hierarchical \textit{enhancement network} to produce $\mathbf{Q}_p \in \mathbb{R}^{C \times F_1 \times T}$, which is then passed through the \textit{FANC decoder} to reconstruct the deterministic spectral estimate $\hat{\mathbf{X}}_{r,i} \in \mathbb{R}^{2 \times F \times T}$. Simultaneously, in the diffusion branch, the noise-perturbed magnitude $\mathbf{X}_t \in \mathbb{R}^{1 \times F \times T}$ is encoded into $\mathbf{E}_o \in \mathbb{R}^{C \times F_1 \times T}$ via the same FANC encoder. To leverage predictive guidance, an initial interaction module harmonizes $\mathbf{E}_o$ with $\mathbf{E}_p$ to produce a synergistic representation $\mathbf{E}_d \in \mathbb{R}^{C \times F \times T}$. This latent is then refined by the hierarchical enhancement network, where cross-branch interaction is performed at every level to inject stable deterministic priors into the diffusion flow. The resulting diffusion features $\mathbf{Q}_d$ are further calibrated with $\mathbf{Q}_p$ to yield $\mathbf{Q}_o \in \mathbb{R}^{C \times F_1 \times T}$, which is ultimately projected through the FANC decoder to estimate the score function $\mathbf{S}_\theta \in \mathbb{R}^{1 \times F \times T}$.

Upon the estimation of the score function $\mathbf{S}_\theta$, the enhancement process concludes with a synergistic fusion and reconstruction phase. To ensure computational tractability, we execute a partial reverse diffusion process starting from a reduced time step $T_{\text{rs}}$, initialized as $\mathbf{X}_{T_{\text{rs}}} \approx \mu(\bar{\mathbf{X}}^{p}_m, \mathbf{Y}_m, T_{\text{rs}}) + \sigma(T_{\text{rs}})\mathbf{z}$. Throughout this iterative refinement, the TLB strategy (Sec.~\ref{sec:inference_strategy}) is seamlessly integrated into the hierarchical U-Net backbone to adaptively modulate the feature maps. After $N$ sampling steps, the resulting generative magnitude $\hat{\mathbf{X}}^{d}_m$ is fused with the predictive magnitude $\bar{\mathbf{X}}^{p}_m$ to yield the final magnitude spectrum:
\begin{equation}
    \hat{\mathbf{X}}_m = \alpha \bar{\mathbf{X}}^{p}_m + (1 - \alpha) \hat{\mathbf{X}}^{d}_m,
\end{equation}
where $\alpha$ serves to balance deterministic stability with generative realism. The final complex spectrogram is then reconstructed by coupling $\hat{\mathbf{X}}_m$ with the phase spectrum inherited directly from the predictive branch, ultimately facilitating high-fidelity speech restoration with minimal computational overhead. A comprehensive summary of DVPD inference procedure is detailed in \textbf{Algorithm~\ref{alg:dvpd_inference}} (see Appendix).

\subsection{Detailed Architectural Components}
\label{Detailed Architectural Components}
The core architecture of DVPD contains: FANC encoder, enhancement network, Frequency-aware Interaction (FI) Module, FANC decoder, which is described as follows:

%Unlike spatially homogeneous natural images, speech spectrograms exhibit non-uniform information density along the frequency axis $F$. To exploit this, we propose the FANC Encoder (Fig.~\ref{f:fig2}(B)). 

%\paragraph{FANC Encoder} For an input $\mathbf{Y} \in \mathbb{R}^{3 \times F \times T}$, FANC (Fig.~\ref{f:fig2}(B)) implements a band-specific partitioning strategy \cite{schroter2022deepfilternet, yu2023efficient}. Unlike spatially homogeneous natural images, speech spectrograms exhibit non-uniform information density along the frequency axis $F$. 1) {Low-band (0--2~kHz)}  This region contains the fundamental frequency $(f0)$ and primary formants, which is preserved without compression to maintain critical harmonic integrity; 2) {Mid-band (2--4~kHz)} undergoes moderate compression; and 3) {High-band ($>4$~kHz)} is heavily compressed to prune spectral redundancy. In contrast to BSRNN \cite{yu2023efficient} or PGUSE \cite{zhang2025composite}, which often over-compress the low-frequency manifold, FANC prioritizes the preservation of the ``acoustic image.'' We employ {heterogeneous dilated kernels} ($3\times3$, $3\times5$, and $3\times7$) to create an \textit{anisotropic receptive field} targeting vertical spectral transients and horizontal harmonics. For the diffusion branch, sinusoidal time embeddings \cite{song2020score, tancik2020fourier} are integrated post fusion. 

\paragraph{FANC Encoder} The FANC encoder (Fig.~\ref{f:fig2}(B)) is designed to exploit the non-uniform information density of speech spectrograms, which distinguishes them from spatially homogeneous natural images. For inputs $\mathbf{Y}\in \mathbb{R}^{3 \times F \times T}$ or  $\mathbf{X}_m \in \mathbb{R}^{1 \times F \times T}$, FANC implements a band-specific partitioning strategy: (\textit{i}) {low-band} ($0$--$2$~kHz), containing the fundamental frequency ($f_0$) and primary formants, is preserved without compression to maintain critical harmonic integrity; (\textit{ii}) {mid-band} ($2$--$4$~kHz) undergoes moderate compression; and (\textit{iii}) {high-band} ($>4$~kHz) is heavily compressed to prune spectral redundancy. In contrast to architectures like BSRNN \cite{yu2023efficient} or PGUSE \cite{zhang2025composite} that often over-compress the low-frequency manifold, FANC prioritizes the preservation of the ``acoustic image.'' This is achieved through {heterogeneous dilated kernels} ($3\times3, 3\times5, \text{and } 3\times7$) that create an \textit{anisotropic receptive field} specifically targeting vertical transients and horizontal harmonics. For the predictive branch, this process yields the encoded features $\mathbf{E}_p \in \mathbb{R}^{C \times F_1 \times T}$. For the diffusion branch, sinusoidal time embeddings \cite{song2020score} are integrated post-fusion, resulting in the generative encoded features $\mathbf{E}_o \in \mathbb{R}^{C \times F_1 \times T}$.

\paragraph{Enhancement Network} To extract hierarchical acoustic features, we employ a 3-layer symmetric U-Net (Fig.~\ref{f:fig2}(C)) as the core enhancement network. Each layer of the backbone is conceptualized as a global-to-local refinement unit, integrating the {Omni-Directional Attention Mechanism (ODAM)} \cite{ke2026omni} and the {LISA} module. Specifically, the predictive features $\mathbf{E}_p$ and diffusion features $\mathbf{E}_d$ are processed independently within each level before undergoing cross-branch calibration via the FI module. We utilize \textit{Pixel Unshuffle} for downsampling and \textit{Pixel Shuffle} for upsampling, effectively mitigating the information loss inherent in traditional pooling. At each level, downsampled features are fused with their counterparts through additive residuals to serve as input for the subsequent stage, ultimately yielding the refined hierarchical representations $\mathbf{Q}_p$ and $\mathbf{Q}_d$.

Taking the first predictive level as an exemplar, the feature $\mathbf{E}_{p}$ is first processed by the ODAM to capture global dependencies and model long-range correlations across both time and frequency axes. The resulting globally-aware feature $\mathbf{E}_{p,1,m}$ is then fed into the {LISA module} to refine spectral textures through a three-stage dynamic filtering process: 
(\textit{i}) \textit{Dynamic Kernel Generation}: instance-specific weights $\mathbf{W}_{d} = \tanh( \text{Conv}_{1 \times 1}(\text{GAP}(\mathbf{E}_{p,1,m})))$ are derived from the global context $\mathbf{E}_{p,1,m}$; 
(\textit{ii}) \textit{Stripe Dynamic Convolution}: anisotropic features are aggregated via $\mathbf{L(\cdot)} = \sum_{k=1}^{K} \mathbf{W}_{d}^{(k)} \odot \mathcal{U}(\text{Pad}(\mathbf{\cdot}), d)^{(k)}$, where $\mathcal{U}(\cdot, d)$ denotes an unfold operation; and 
(\textit{iii}) \textit{Dual-path Refinement}: features are fused to balance structural stability and detail:
\begin{equation}
\begin{aligned}
    \mathbf{O}_{F,d} &= \lambda_{F1,d} \odot \mathbf{L(\mathbf{E}_{p,1,m})} + \lambda_{F2,d} \odot \mathbf{E}_{p,1,m}, \\
    \mathbf{N}_{d} &= \lambda_{T1,d} \odot \mathbf{L(\mathbf{O}_{F,d})} + \lambda_{T2,d} \odot \mathbf{O}_{F,d}.
\end{aligned}
\end{equation}
The final refined output $\mathbf{E}_{p,1}$ is formulated as $\mathbf{E}_{p,1} = \text{Conv} ( \sum_{d \in \{3, 5, 7\}} (\gamma_d \Psi_d(\mathbf{E}_{p,1,m}) + \beta_d \mathbf{N}_d) )$, where $\Psi_d(\cdot)$ represents sequential $\mathcal{T}$ and $\mathcal{F}$ stripe operations with multi-scale dilations $d \in \{3, 5, 7\}$. 
\paragraph{Frequency-aware Interaction (FI) Module} The FI module (Fig.~\ref{f:fig2}(D)) adaptively calibrates the cross-branch feature exchange through a dual-path gating mechanism. Given the diffusion feature $\mathbf{E}_{d,i}$, the predictive feature $\mathbf{E}_{p,i}$, and the time embedding $t_{emb}$, the interaction unfolds as follows: (\textit{i}) \textit{Joint Feature Fusion}: $\mathbf{E}_{d,i}$ and $\mathbf{E}_{p,i}$ are concatenated and processed via a \textit{Conv2D} layer, with $t_{emb}$ added to inject temporal-diffusion context; (\textit{ii}) \textit{Global Prior Extraction}: a parallel path applies \textit{AvgPool2d} and \textit{Conv2D} to $\mathbf{E}_{p,i}$ to capture global spectral-temporal dependencies. Both paths are then passed through \textit{Sigmoid} functions to generate dynamic importance masks, which are element-wise multiplied to produce a unified reliability weight. This weight modulates the predictive guidance $\mathbf{E}_{p,i}$ before it is residually added to the original diffusion latent. This process yields the final interaction output $\mathbf{E}_{out,i}$, ensuring that the generative process prioritizes robust low-frequency deterministic priors while effectively suppressing unreliable, noise-dominated regions in the predictive cues.

\paragraph{FANC Decoder} The FANC decoder symmetrically mirrors the partitioning structure of the encoder to facilitate resolution recovery. Taking the hierarchical features $\mathbf{Q}_p$ and $\mathbf{Q}_o$ as inputs, the decoder utilizes sub-pixel convolutions (SP-Conv2D) \cite{shi2016real} to inversely map the compressed latent representations back to the original spectral resolution $F \times T$. This process ultimately yields the deterministic complex estimate $\hat{\mathbf{X}}_{r,i} \in \mathbb{R}^{2 \times F \times T}$ from the predictive branch and the score estimate $\mathbf{S}_\theta \in \mathbb{R}^{1 \times F \times T}$ from the diffusion branch.

\subsection{TLB strategy}
\label{sec:inference_strategy}

\begin{figure}[htb]
\begin{minipage}[b]{1.0 \linewidth}
 \centering
 % \hspace{3cm}  % 向you移动7厘米
 \centerline{\includegraphics[width= \columnwidth]{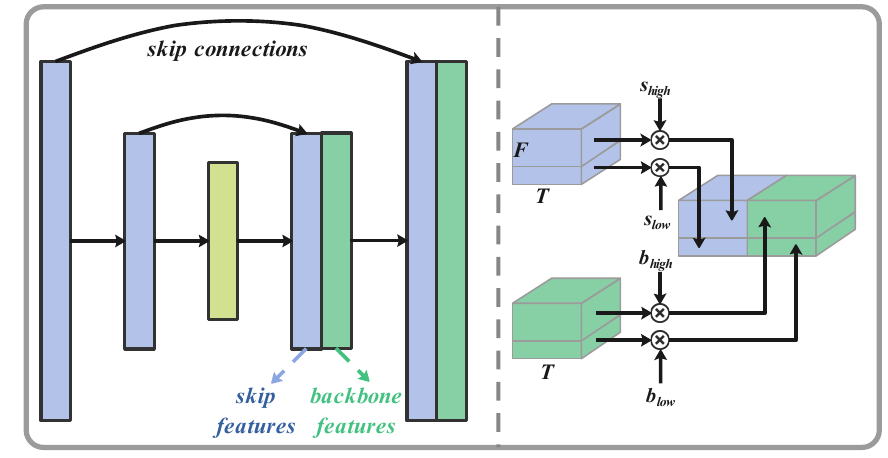}}
\end{minipage}

% \vspace{-0.5cm}
\caption{Operations of the TLB strategy. Scaling factors $b$ and $s$ modulate the intensity of backbone and skip features, respectively.}

\label{f:fig4}

\end{figure}

\paragraph{TLB strategy}
 Treating the spectrogram as a structural 2D manifold allows us to analyze the denoising dynamics within the U-Net backbone. Recent studies \cite{si2024freeu} suggest that the U-Net backbone inherently prioritizes low-frequency structural restoration while aggressively suppressing high-frequency noise. Conversely, skip connections are instrumental in recovering fine-grained transients but may inadvertently propagate residual noise or expedite premature convergence during the generative process.
Building on our \textit{dual-view} perspective, we propose the TLB strategy, an inference technique that modulates feature maps without additional training (Fig.~\ref{f:fig4}). We define four scaling factors: $b_{{low}}, b_{{high}}$ (backbone) and $s_{{low}}, s_{{high}}$ (skip). The $b$ factors are formulated to modulate the backbone's high-frequency suppression intensity, whereas the $s$ factors recalibrate the spectral energy integrity of low-frequency harmonics within the skip connections. By aligning these factors with the spectrogram's anisotropic properties, where ``low" and ``high"  (Fig.~\ref{f:fig1}(left)) correspond to the distinct harmonic and transient regions of the spectrogram, TLB adaptively balances noise reduction and fine-grained detail preservation during the generative refinement stage. Please note that we only use the TLB strategy on the diffusion branch. For a more detailed description of using the spectrum as different frequency components for image denoising, see \textbf{Appendix~\ref{app:TLB_philosophy}}.

\begin{table*}[t]
\centering
\caption{Speech enhancement results where all models are {both trained and evaluated on the WSJ0-UNI}. \textbf{Bolds} indicate the best while \underline{underlines} indicate the second best. MACs represent the total computational complexity of the entire inference process per second of audio.} % Notably, all results in this table are reported \textbf{without} TLB strategy to establish a fair baseline comparison.}
\label{tab:1}
\resizebox{\textwidth}{!}{%
\begin{tabular}{lccccccccc} % 调整为10列
\toprule
Method & Para. & MACs & Type & PESQ $\uparrow$ & ESTOI $\uparrow$ & CSIG $\uparrow$ & CBAK $\uparrow$ & COVL $\uparrow$ & WV-MOS $\uparrow$ \\ \midrule
Degraded & - & - & - & 1.67 $\pm$ 0.60 & 0.70 $\pm$ 0.18 & 2.41 $\pm$ 1.15 & 1.92 $\pm$ 0.60 & 2.01 $\pm$ 0.87 & 1.79 $\pm$ 2.13 \\ \midrule

Conv-TasNet \cite{luo2019conv} & 3.4M & \underline{3.2G} & P & 2.01 $\pm$ 1.21 & 0.76 $\pm$ 0.09 & 3.02 $\pm$ 0.99 & 2.23 $\pm$ 0.78 & 2.64 $\pm$ 1.01 & 2.56 $\pm$ 1.33 \\
MANNER \cite{park2022manner} & 24.1M & 8.7G & P & 2.21 $\pm$ 1.23 & 0.80 $\pm$ 0.05 & 3.41 $\pm$ 0.66 & 2.46 $\pm$ 0.91 & 2.78 $\pm$ 1.03 & 2.99 $\pm$ 0.55 \\
PGUSE-P \cite{zhang2025composite} & 2.3M & 5.8G & P & 2.38 $\pm$ 1.10 & \underline{0.85 $\pm$ 0.11} & 3.43 $\pm$ 0.73 & 2.60 $\pm$ 0.51 & 2.91 $\pm$ 0.61 & 3.21 $\pm$ 0.93 \\
CMGAN \cite{abdulatif2024cmgan} & \underline{1.8M} & 31.7G & P & 2.66 $\pm$ 0.99 & \textbf{0.88 $\pm$ 0.09} & 3.52 $\pm$ 0.81 & 2.81 $\pm$ 0.85 & 3.11 $\pm$ 0.62 & 3.55 $\pm$ 0.55 \\
MP-SENet \cite{lu2023mp} & 2.26M & 34.58G & P & \textbf{2.71 $\pm$ 0.89} & \textbf{0.88 $\pm$ 0.13} & \textbf{3.99 $\pm$ 0.76} & \underline{2.90 $\pm$ 0.58} & \textbf{3.38 $\pm$ 0.89} & \textbf{4.16 $\pm$ 0.25} \\ \midrule

{DVPD-P (\textit{ours})} & \textbf{0.61M} & \textbf{2.41G} & P & \underline{2.70 $\pm$ 0.99} & \textbf{0.88 $\pm$ 0.08} & \underline{3.91 $\pm$ 0.88} & \textbf{2.91 $\pm$ 0.59} & \underline{3.28 $\pm$ 0.99} & \underline{3.76 $\pm$ 0.47} \\ 
 \midrule

CDiffuSE \cite{lu2022conditional} & \underline{4.3M} & 292.4G & D & 1.97 $\pm$ 0.91 & 0.80 $\pm$ 0.11 & 2.77 $\pm$ 0.71 & 1.99 $\pm$ 0.92 & 2.21 $\pm$ 1.01 & 2.50 $\pm$ 1.05 \\
SGMSE+ \cite{richter2023speech} & 65.6M & 8.0T & D & 2.61 $\pm$ 1.12 & 0.90 $\pm$ 0.11 & 3.79 $\pm$ 0.85 & 2.65 $\pm$ 0.81 & 3.09 $\pm$ 1.15 & 3.40 $\pm$ 0.91 \\
DOSE+ \cite{yang2025dose+} & 65.9M & 310.4G & D & 2.84 $\pm$ 0.66 & {0.90 $\pm$ 0.10} & 3.96 $\pm$ 0.63 & 2.79 $\pm$ 0.41 & {3.49 $\pm$ 0.79} & {3.69 $\pm$ 0.71} \\
StoRM \cite{lemercier2023storm} & 55.1M & 15.8T & D+P & 2.75 $\pm$ 1.03 & 0.89 $\pm$ 0.08 & 3.84 $\pm$ 0.77 & 2.66 $\pm$ 0.41 & 3.09 $\pm$ 0.90 & 3.44 $\pm$ 0.78 \\
UNIVERSE++ \cite{scheibler2024universal} & 42.9M & 42.8G & D+P & 2.66 $\pm$ 0.93 & 0.89 $\pm$ 0.11 & 3.89 $\pm$ 0.59 & 2.60 $\pm$ 0.49 & 3.21 $\pm$ 0.88 & 3.46 $\pm$ 0.77 \\

PGUSE \cite{zhang2025composite} & 5.1M & \underline{26.3G} & D+P & 2.95 $\pm$ 0.91 & \underline{0.91 $\pm$ 0.06} & 4.01 $\pm$ 0.77 & 2.61 $\pm$ 0.60 &\textbf{3.53 $\pm$ 0.91} & {3.44 $\pm$ 0.66} \\ \midrule

{DVPD (\textit{ours}) (\textit{without TLB})} & \textbf{1.9M} & \textbf{10.2G} & D+P & \underline{2.99 $\pm$ 0.88} & \underline{0.91 $\pm$ 0.12} & \underline{4.06 $\pm$ 0.71} & \underline{2.93 $\pm$ 0.57} & {3.43 $\pm$ 0.87} & \underline{4.16 $\pm$ 0.25} \\
{DVPD (\textit{ours}) (\textit{with TLB)}} & \textbf{1.9M} & \textbf{10.2G} & D+P & \textbf{3.15 $\pm$ 0.79} & \textbf{0.92 $\pm$ 0.05} & \textbf{4.21 $\pm$ 0.37} & \textbf{3.01 $\pm$ 0.47} & \underline{3.51 $\pm$ 0.99} & \textbf{4.27 $\pm$ 0.31} \\
\bottomrule

\end{tabular}%
}
\end{table*}

\section{Experiments}

\subsection{Datasets}

To evaluate the efficacy and versatility of our model across diverse acoustic conditions, we conduct extensive experiments on several widely recognized benchmarks: We utilize the {WSJ0-UNI} \cite{ristea2025icassp} dataset, which encompasses a broad range of distortions, to benchmark the model's performance on Universal Speech Enhancement (USE). The {VoiceBank+DEMAND} (VBDMD) \cite{botinhao2016investigating} dataset is employed as a standard benchmark for evaluating localized noise suppression capabilities. {VBDMD-SR} for evaluating speech super-resolution. To further assess cross-task robustness, we evaluate our model on {VBDMD-REVERB}(VBD-RB), {WSJ0-REVERB}(WSJ0-RB) \cite{garofolo2007csr} and {WSJ0-CHiME3}(WSJ0-CE3) \cite{barker2015third} to verify the model's out-of-distribution generalization in real world noisy and reverberant environments. All audio samples are resampled to 16\,kHz for consistency. See \textbf{Appendix~\ref{app:datasets}} for more details.

\subsection{Model Configurations and Evaluation Metrics}
\label{sec:implementation}

All audio signals were processed in the T-F domain using an STFT with a 512-point Hann window and a 128-point hop size. The U-Net backbone employed encoder/decoder stages with an initial channel 24, doubling at each level to 96 bottleneck. Each stage consisted of a single module, except for the diffusion bottleneck, which comprised two. For the BBED SDE, we set $k=2.6$, $c=0.51$, $T=0.999$ \cite{lay2023reducing}, $\alpha=0.4$, $T_{\text{rs}}=0.12$ and $N$ = 3 (\textbf{Appendix~\ref{app:inference_ablation}}). The model was trained for 200 epochs using the AdamW on 4 NVIDIA RTX A6000 GPUs with a batch size of 32. The learning rate was $1 \times 10^{-3}$, decaying by 0.97 every two epochs, with gradient clipping at an $L_2$ norm of 5.0.
The total loss $\mathcal{L}$ was a combination of predictive and generative objectives:
\begin{equation}
    \mathcal{L} = \lambda_1 \mathcal{L}_{\text{mag}} + (1 - \lambda_1) \mathcal{L}_{\text{comp}} + \lambda_2 \mathcal{L}_{\text{pha}} + \mathcal{L}_{\text{score}},
\end{equation}
where $\mathcal{L}_{\text{mag}}$, $\mathcal{L}_{\text{comp}}$, and $\mathcal{L}_{\text{pha}}$ denoted the magnitude, complex spectral, and anti-wrapping phase losses for the predictive branch, respectively, and $\mathcal{L}_{\text{score}}$ was the score-matching loss. We empirically set $\lambda_1 = 0.5$ and $\lambda_2 = 0.002$ with many attempts to ensure that all loss components remained within the same order of magnitude during training.

Performance was evaluated using a comprehensive suite of metrics, including {PESQ} \cite{rix2001perceptual}, {ESTOI} \cite{jensen2016algorithm}, {SI-SDR} \cite{le2019sdr}, {WV-MOS} \cite{andreev2023hifi++}, DNS-MOS \cite{reddy2021dnsmos} and the composite Mean Opinion Score (MOS) estimates ({CSIG}, {CBAK}, {COVL}) \cite{hu2007evaluation}. To demonstrate efficiency, we also reported the number of {Parameters (M)} and {MACs} \footnote{\url{https://github.com/sovrasov/flops-counter.pytorch}}. More Details were provided in \textbf{Appendix \ref{app:loss_and_metrics}}.

\begin{figure*}[h]
\begin{minipage}[b]{1.0 \linewidth}
 \centering
 % \hspace{3cm}  % 向you移动7厘米
 \centerline{\includegraphics[width= \columnwidth]{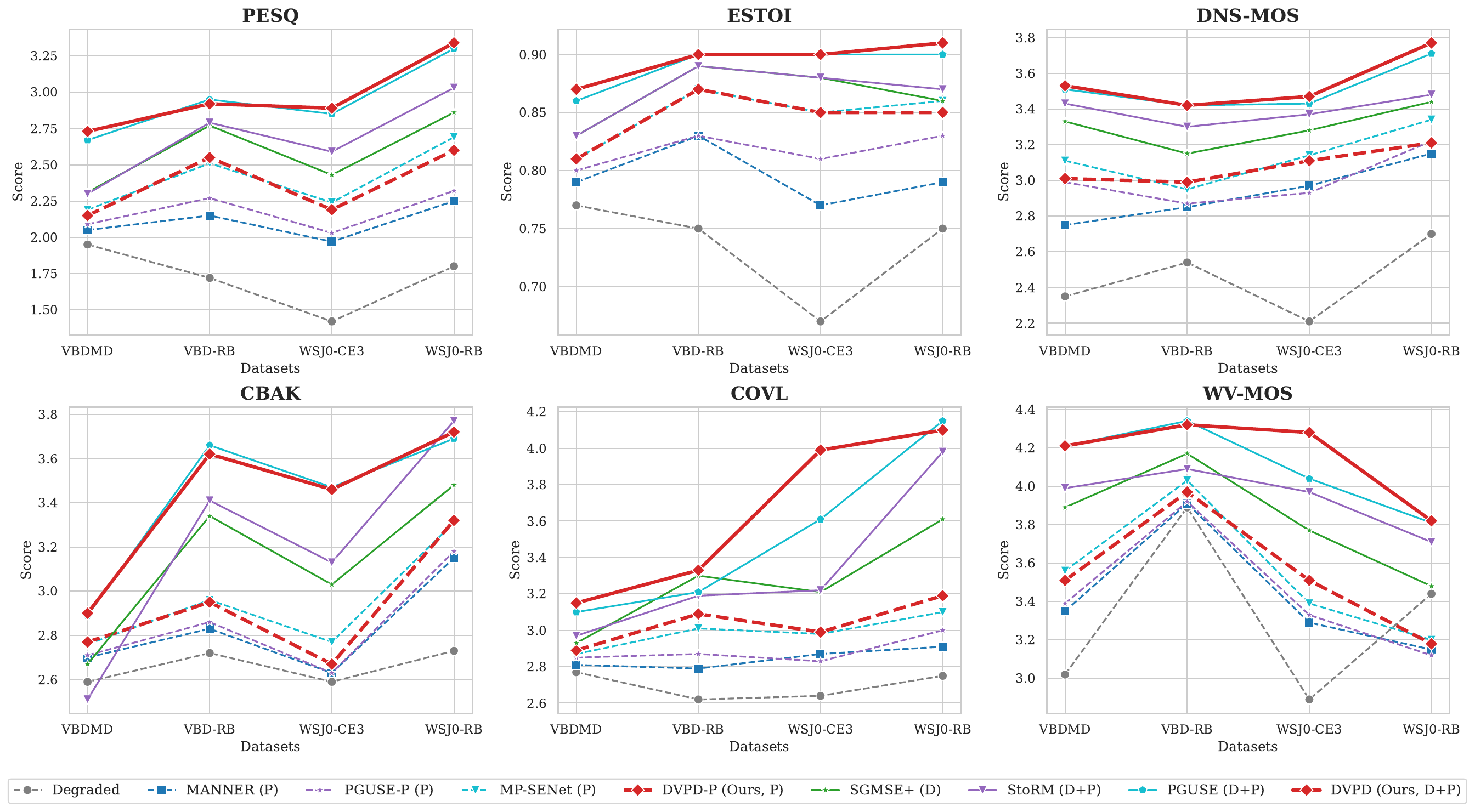}}
\end{minipage}

% \vspace{-0.5cm}
\caption{Performance comparison across diverse benchmarks. All models were trained exclusively on the WSJ0-UNI to evaluate their universal generalization capabilities. Dashed lines represent predictive models, while solid lines denote diffusion-based generative models.}

\label{f:figwsj0}

\end{figure*}

\subsection{Comparisons with Methods on USE}
\label{Comparisons with Methods on USE}

As demonstrated in Table~\ref{tab:1}, we evaluated the USE performance of our proposed model on the WSJ0-UNI dataset, comparing against various SOTA predictive and generative baselines in terms of parameter, computational complexity (MACs), and speech quality metrics. By decoupling our framework into its predictive and generative components, we observe that the predictive branch, {DVPD-P}, achieves speech quality comparable to the leading MP-SENet while utilizing only one-third of its parameters and reducing MACs by an order of magnitude. In the diffusion domain, the results confirm that hybrid predictive-diffusion (D+P) frameworks consistently outperform pure diffusion (D) models; notably, compared to the previous SOTA lightweight model PGUSE, our DVPD requires only {35\%} of the parameters and {40\%} of the inference MACs while delivering superior performance across most objective measures. Furthermore, by incorporating our TLB strategy during inference, the performance of DVPD is significantly enhanced without any additional training or computational overhead, further validating the efficiency and robustness of our dual-view architectural design.

\subsection{Evaluation of Out-of-Distribution Generalization}
\label{sec:ood_generalization}

%To verify the robustness of our model under distributional shifts, Figure~\ref{f:figwsj0} illustrates the {zero-shot generalization} performance of various models trained exclusively on WSJ0-UNI. Specifically, the {VBDMD} and {VBD-RB} datasets assess the model's adaptability to unseen clean speech sources, noise types, and reverberation. In contrast, {WSJ0-CE3} and {WSJ0-RB} evaluate the compatibility of the models when the clean source remains the same but is subjected to entirely distinct noise and reverberant profiles. 

%Several key observations emerge. First, diffusion-based models (solid lines) consistently exhibit superior generalization across nearly all benchmarks compared to predictive models (dashed lines), highlighting the {inherent advantage of stochastic refinement} in handling unseen distortions. Within the predictive category, our {DVPD}-P achieves performance comparable to the SOTA MP-SENet, even surpassing it in specific scenarios. This is particularly noteworthy given that {DVPD}-P possesses significantly {fewer parameters and lower computational complexity}. For the generative paradigm, our full {DVPD} model sets a new SOTA benchmark across the majority of datasets, further validating the efficacy of our dual-view architectural design in bridging the gap between efficiency and universal robustness.
To evaluate robustness against distributional shifts, we assessed the zero-shot generalization of models trained exclusively on WSJ0-UNI (Fig.~\ref{f:figwsj0}) across unseen speech sources, noise types, and reverberant profiles (VBDMD, VBD-RB, WSJ0-CE3, and WSJ0-RB). Several key observations emerge. First, diffusion-based models (solid lines) consistently exhibit superior generalization across nearly all benchmarks compared to predictive models (dashed lines), highlighting the {inherent advantage of stochastic refinement} in handling unseen distortions. Within the predictive category, our {DVPD}-P achieves performance comparable to the SOTA MP-SENet, even surpassing it in specific scenarios. This is particularly noteworthy given that {DVPD}-P possesses significantly {fewer parameters and lower computational complexity}. For the generative paradigm, our full {DVPD} model sets a new SOTA benchmark across the majority of datasets, further validating the efficacy of our dual-view architectural design in bridging the gap between efficiency and universal robustness.
\subsection{Speech Denoising on VBDMD}
\label{sec:vbdmd_results}

Table~\ref{tab:speech_enhancement_vbdmd_results} presents the performance of our model trained and evaluated specifically on the VBDMD benchmark. In this single distortion (noise-only) scenario, the spectral mapping task is relatively less complex compared to universal speech enhancement. Consequently, we observe that predictive models generally exhibit a performance edge over generative ones. Specifically, {MP-SENet} achieves the highest scores; this is likely because, in the absence of catastrophic information loss, the {mean-regression} characteristic of predictive models is highly effective at capturing the deterministic mapping between noisy and clean spectral manifolds. 

\begin{table}[h]
\centering
\caption{Performance comparison on the VBDMD dataset. All models are {trained on the VBDMD}. The symbol $\dagger$ denotes results obtained from our own implementations and evaluations.}
\label{tab:speech_enhancement_vbdmd_results}
\resizebox{\columnwidth}{!}{%
\begin{tabular}{lcccc}
\toprule
Method & PESQ $\uparrow$ & ESTOI $\uparrow$ & CSIG $\uparrow$ & DNS-MOS $\uparrow$ \\ \midrule
Degraded & 1.98 & 0.79 & 3.48 & 2.39 \\ \midrule
% \textit{Predictive Methods} \\
Conv-TasNet \cite{luo2019conv} & 2.56 & 0.85 & 3.89 & 3.33 \\
PGUSE-P$\dagger$ \cite{zhang2025composite} & 3.09 & 0.87 & \underline{4.45} & \underline{3.59} \\

MP-SENet \cite{lu2023mp} & \textbf{3.50} & \textbf{0.91} & \textbf{4.73} & \textbf{3.67} \\\midrule 
{DVPD-P (\textit{ours})} & \underline{3.14} & \underline{0.88} & 4.44 & \underline{3.59} \\
 \midrule
% \textit{Generative Methods} \\
CDiffuSE \cite{lu2022conditional} & 2.48 & 0.79 & 3.77 & 3.31 \\
SGMSE+ \cite{richter2023speech} & 2.88 & 0.86 & 4.24 & 3.45 \\
StoRM \cite{lemercier2023storm} & 2.85 & 0.87 & 4.18 & 3.43 \\
UNIVERSE++ \cite{scheibler2024universal} & 3.03 & 0.87 & 4.38 & 3.51 \\ 
PGUSE$\dagger$ \cite{zhang2025composite} & {3.11} & \underline{0.88} & {4.61} & \underline{3.58} \\
FlowSE \cite{lee2025flowse} & \underline{3.12} & \underline{0.88} & \underline{4.62} & \underline{3.58} \\ \midrule  % 3.58
% \textbf{DVPD(Ours)} & \textbf{3.27} & \textbf{0.88} & \textbf{4.67} & \textbf{4.35} \\
{DVPD (\textit{ours})} & \textbf{3.35} & \textbf{0.89} & \textbf{4.73} & \textbf{3.66} \\
\bottomrule
\end{tabular}%
}
\end{table}

Within this context, our predictive branch, {DVPD-P}, delivers performance second only to MP-SENet. More importantly, while pure generative models typically lag behind SOTA predictive models in localized denoising tasks, our DVPD (D+P) significantly narrows the performance gap between the two paradigms. By integrating the stable deterministic priors from the predictive branch, DVPD demonstrates exceptional versatility, proving its effectiveness even in simplified, single modality denoising scenarios.

\begin{table}[h]
\centering
\caption{Performance comparison on the VBDMD-SR dataset (8kHz - 16kHz). Models are trained on WSJ0-UNI.}
\label{tab:speech_enhancement_results}
\resizebox{\columnwidth}{!}{%
\begin{tabular}{lcccc}
\toprule
Method & PESQ $\uparrow$ & ESTOI $\uparrow$ & COVL $\uparrow$ & CSIG $\uparrow$ \\ \midrule
Degraded & 4.22 & 0.95 & 2.99 & 1.69 \\ \midrule
% \textit{Predictive Methods} \\
Conv-TasNet \cite{luo2019conv} & 3.48 & 0.91 & 3.89 & 4.21 \\

MP-SENet \cite{lu2023mp} & \textbf{3.79} & \underline{0.90} & \underline{4.09} & \textbf{4.56} \\\midrule 
{DVPD-P (\textit{ours})} & \underline{3.70} & \textbf{0.91} & \textbf{4.15} & \underline{4.39} \\
 \midrule
% \textit{Generative Methods} \\
CDiffuSE \cite{lu2022conditional} & 2.66 & 0.86 & 3.08 & 3.42 \\
SGMSE+ \cite{richter2023speech} & 3.84 & 0.92 & 3.91 & 3.86 \\
StoRM \cite{lemercier2023storm} & 2.89 & 0.88 & 3.24 & 3.50 \\
UNIVERSE++ \cite{scheibler2024universal} & 3.01 & 0.87 & 3.52 & 3.93 \\ 
PGUSE \cite{zhang2025composite} & \underline{4.09} & \underline{0.94} & \textbf{4.32} & \underline{4.41} \\ \midrule
{DVPD (\textit{ours})} & \textbf{4.15} & \textbf{0.96} & \underline{4.28} & \textbf{4.44} \\ \bottomrule
\end{tabular}%
}
\end{table}

\subsection{Evaluation on Speech Super-Resolution}
\label{sec:sr_results}

We further evaluate the performance of DVPD on the {VBDMD-SR} dataset using the model trained on WSJ0-UNI. Speech super-resolution (SR), also known as bandwidth extension, is a widespread yet fundamentally {ill-posed generative task} that requires reconstructing high-frequency spectral content from band-limited observations. Compared to other baseline methods, DVPD maintains a leading position in most evaluation metrics, demonstrating its superior generative capacity.
An interesting observation is that band-limited degraded speech achieves the highest PESQ score, while all enhanced models exhibit a performance decline in this specific metric. This phenomenon arises because the PESQ algorithm is not explicitly designed for SR evaluation, it is heavily biased toward low-frequency spectral fidelity, which dominates human auditory perception. During the generation process, models inevitably introduce minor reconstructive errors in the low-frequency regions, leading to a penalty in PESQ scores despite the successful reconstruction of high-frequency components. 
% To mitigate this, SE methods for SR typically employ \textbf{Low-Frequency Replacement (LFR)} \cite{liu2023neural}, which reuses the original low-band components from the input signal to ensure spectral integrity. Upon integrating the LFR strategy, the performance of DVPD is substantially improved, achieving a significant boost in both perceptual quality and signal fidelity.

\begin{table}[t]
\centering
\caption{Ablation study on the WSJ0-UNI, evaluating the impact of key components and SDE formalisms. All results demonstrate the contribution of each module to the overall performance.}
\label{table:ablation}

\resizebox{\columnwidth}{!}{%
\begin{tabular}{lcccc}
\toprule
Configuration & PESQ & CSIG & CBAK & WV-MOS \\
\midrule
{(DVPD)} & \textbf{2.99} & \textbf{4.06} & \textbf{2.93} & \textbf{4.16} \\
\midrule

w/o FANC Encoder & \underline{2.93} & 3.98 & 2.86 & 4.09  \\
w/o FI Module & 2.91 & \underline{4.01} & \underline{2.88} & \underline{4.12} \\
w/o Phase Loss & 2.91 & 3.99 & \underline{2.88} & 4.11  \\
w/o LISA Module & 2.71 & 3.79 & 2.75 & 3.85  \\
\midrule
w/ OUVE SDE & 2.89 & 3.95 & 2.81 & 4.06 \\
w/ Degraded Phase & 2.65 & 3.66 & 2.61 & 3.71 \\
\bottomrule
\end{tabular}
}
\end{table}

\subsection{Ablation Study}
\label{sec:ablation_study}

\paragraph{Contribution of key components}
To quantify the contribution of each key component within the DVPD framework, we conducted a comprehensive ablation study on the WSJ0-UNI dataset, as summarized in Table~\ref{table:ablation}. Detailed configuration variants for each ``without" (w/o) case are provided in \textbf{Appendix~\ref{app:ablation_config}}. From the architectural perspective, removing the {FANC Encoder}, the {FI Module}, or the {Phase Loss} leads to a measurable decrease in performance, although the impact is relatively moderate. However, the exclusion of the {LISA Module} results in a significant performance degradation across all perceptual metrics. This pronounced decline underscores the critical importance of our multi-range processing approach implemented via the LISA module. 
Regarding the diffusion formalism and phase strategy, replacing the {BBED} formulation with {OUVE} negatively impacts the efficacy of the model, which can be attributed to the inherent prior mismatch issue associated with the OUVE SDE. Furthermore, we observe that substituting our predicted phase with the original degraded phase for final waveform reconstruction leads to a substantial drop in speech quality. This validates our strategy of leveraging the predictive branch for high-fidelity phase estimation to guide the generative process.

\begin{table}[t]
\caption{Performance gain of the {TLB strategy} across different quality tiers on WSJ0-UNI and VBDMD benchmarks. Tiers are categorized by the baseline PESQ scores. All results are averaged over 20 independent inference runs to ensure statistical stability.}
\label{table:TLB_tiers_analysis}
\vskip 0.15in
\begin{center}
\resizebox{\columnwidth}{!}{
\begin{tabular}{llccccc}
\toprule
Dataset & Performance Tier & Method & PESQ $\uparrow$ & ESTOI $\uparrow$ & SI-SDR & WV-MOS $\uparrow$ \\
\midrule
\multirow{6}{*}{WSJ0-UNI} & \multirow{2}{*}{(PESQ $<$ 2)} & Base & 1.73 & 0.75 & \textbf{16.99} & 3.12 \\
 & & \textbf{+ TLB} & \textbf{1.95} (\textit{+0.22}) & \textbf{0.78} & 15.85 & \textbf{3.35} \\
 \cmidrule{2-7}
 & \multirow{2}{*}{(2 $\le$ PESQ $<$ 3)} & Base & 2.45 & 0.84 & \textbf{18.50} & 3.85 \\
 & & \textbf{+ TLB} & \textbf{2.51} (\textit{+0.06}) & \textbf{0.85} & 18.42 & \textbf{3.92} \\
 \cmidrule{2-7}
 & \multirow{2}{*}{(PESQ $\ge$ 3)} & Base & 3.20 & 0.92 & \textbf{19.30} & 4.45 \\
 & & \textbf{+ TLB} & \textbf{3.24} (\textit{+0.04}) & \textbf{0.92} & 19.25 & \textbf{4.46} \\
\midrule
\multirow{6}{*}{VBDMD} & \multirow{2}{*}{(PESQ $<$ 2)} & Base & 1.73 & 0.77 & \textbf{13.99} & 2.23 \\
 & & \textbf{+ TLB} & \textbf{1.88} (\textit{+0.15}) & \textbf{0.78} & 12.43 & \textbf{2.41} \\
 \cmidrule{2-7}
 & \multirow{2}{*}{(2 $\le$ PESQ $<$ 3)} & Base & 2.59 & \textbf{0.82} & \textbf{17.29} & 3.35 \\
 & & \textbf{+ TLB} & \textbf{2.65} (\textit{+0.06}) & \textbf{0.82} & 16.59 & \textbf{3.55} \\
 \cmidrule{2-7}
 & \multirow{2}{*}{(PESQ $\ge$ 3)} & Base & 3.56 & 0.92 & \textbf{20.81} & 4.58 \\
 & & \textbf{+ TLB} & \textbf{3.60} (\textit{+0.04}) & \textbf{0.92} & 19.31 & \textbf{4.63} \\
\bottomrule
\end{tabular}
}
\end{center}
\vskip -0.1in
\end{table}

\paragraph{Analysis of the TLB Strategy}
To further investigate the efficacy of the proposed {TLB strategy}, we stratify the test samples from WSJ0-UNI and VBDMD into three distinct quality tiers based on their baseline PESQ scores: (PESQ $<$ 2), (2 $\le$ PESQ $<$ 3), and (PESQ $\ge$ 3). This stratification is motivated by the varying spectro-temporal characteristics across different quality levels. Accordingly, the scaling parameters ($s, b$) are fine-tuned for each tier to achieve optimal restoration; The rationale for quality stratification and the sensitivity of parameters ($s, b$) are further elaborated in \textbf{Appendix~\ref{app:Performance Analysis of TLB}}.
As shown in Table~\ref{table:TLB_tiers_analysis}, the TLB strategy yields the most substantial perceptual gains in the (PESQ $<$ 2) across both datasets. This phenomenon can be attributed to the fact that in severely degraded scenarios, the base model often fails to fully reconstruct the fundamental harmonic manifold in the low-frequency regions. Our TLB strategy effectively compensates for these missing structural cues by recalibrating and injecting salient features from the skip-connections, thereby significantly enhancing the harmonic integrity and perceptual quality.
Notably, the marginal SI-SDR decrease reflects a classic perception-fidelity tradeoff. While modulating skip-connection features enhances perceptual clarity, it inherently introduces minor sample-level variations, prioritizing spectral realism over strict waveform alignment.

\section{Conclusion}
We presented DVPD, a novel predictive-diffusion framework that leverages a dual-view perspective to model spectrograms as both physical frequency representations and visual textures. Through the FANC encoder, LISA module, and the TLB strategy, our model significantly reduces computational overhead while improving restoration quality. Experimental results show that DVPD outperforms much larger SOTA models using only 35\% of the parameters and 40\% of the MACs, which provides a robust solution for real-world SE.
% By bridging acoustic physics with efficient architectural design, DVPD provides a robust and scalable solution for real-world universal speech enhancement.

\section*{Impact Statement}

% Authors are \textbf{required} to include a statement of the potential broader
% impact of their work, including its ethical aspects and future societal
% consequences. This statement should be in an unnumbered section at the end of
% the paper (co-located with Acknowledgements -- the two may appear in either
% order, but both must be before References), and does not count toward the paper
% page limit. In many cases, where the ethical impacts and expected societal
% implications are those that are well established when advancing the field of
% Machine Learning, substantial discussion is not required, and a simple
% statement such as the following will suffice:

This paper presents work whose goal is to advance the field of Machine
Learning. There are many potential societal consequences of our work, none
which we feel must be specifically highlighted here.

% The above statement can be used verbatim in such cases, but we encourage
% authors to think about whether there is content which does warrant further
% discussion, as this statement will be apparent if the paper is later flagged
% for ethics review.

% % In the unusual situation where you want a paper to appear in the
% % references without citing it in the main text, use \nocite

\bibliography{example_paper}
\bibliographystyle{icml2026}

%%%%%%%%%%%%%%%%%%%%%%%%%%%%%%%%%%%%%%%%%%%%%%%%%%%%%%%%%%%%%%%%%%%%%%%%%%%%%%%
%%%%%%%%%%%%%%%%%%%%%%%%%%%%%%%%%%%%%%%%%%%%%%%%%%%%%%%%%%%%%%%%%%%%%%%%%%%%%%%
% APPENDIX
%%%%%%%%%%%%%%%%%%%%%%%%%%%%%%%%%%%%%%%%%%%%%%%%%%%%%%%%%%%%%%%%%%%%%%%%%%%%%%%
%%%%%%%%%%%%%%%%%%%%%%%%%%%%%%%%%%%%%%%%%%%%%%%%%%%%%%%%%%%%%%%%%%%%%%%%%%%%%%%
\newpage
\appendix
\onecolumn

% You can have as much text here as you want. The main body must be at most $8$
% pages long. For the final version, one more page can be added. If you want, you
% can use an appendix like this one.

% The $\mathtt{\backslash onecolumn}$ command above can be kept in place if you
% prefer a one-column appendix, or can be removed if you prefer a two-column
% appendix.  Apart from this possible change, the style (font size, spacing,
% margins, page numbering, etc.) should be kept the same as the main body.
%%%%%%%%%%%%%%%%%%%%%%%%%%%%%%%%%%%%%%%%%%%%%%%%%%%%%%%%%%%%%%%%%%%%%%%%%%%%%%%
%%%%%%%%%%%%%%%%%%%%%%%%%%%%%%%%%%%%%%%%%%%%%%%%%%%%%%%%%%%%%%%%%%%%%%%%%%%%%%%

\appendix
\section{Extended Background on Score-based Diffusion Models}
\label{app:diffusion_details}

\subsection{Forward and Reverse SDEs}
The forward diffusion process $\{ \mathbf{x}_t \}_{t=0}^T$ is defined by the following Itô SDE:
\begin{equation}
    d\mathbf{x}_t = \mathbf{f}(\mathbf{x}_t, t)dt + g(t)d\mathbf{w},
\end{equation}
where $\mathbf{f}(\cdot, t)$ is the drift coefficient, $g(t)$ is the diffusion coefficient, and $\mathbf{w}$ denotes the standard $d$-dimensional Brownian motion. In SE, this process is applied independently to each T-F bin in the STFT domain.

Under certain regularity conditions, the reverse-time process $\{ \mathbf{x}_t \}_{t=T}^0$ satisfies \cite{anderson1982reverse, song2020score}:
\begin{equation}
    d\mathbf{x}_t = [-\mathbf{f}(\mathbf{x}_t, t) + g^2(t) \nabla_{\mathbf{x}_t} \log p_t(\mathbf{x}_t)] dt + g(t)d\bar{\mathbf{w}},
\end{equation}
where $d\bar{\mathbf{w}}$ is the reverse-time Brownian motion. The term $\nabla_{\mathbf{x}_t} \log p_t(\mathbf{x}_t)$ is the score function, which we approximate using a neural network $\mathbf{s}_\theta(\mathbf{x}_t, \mathbf{y}, t)$ trained via denoising score matching:
\begin{equation}
    \mathcal{L}_{\text{score}} = \mathbb{E}_{t, \mathbf{x}_0, \mathbf{y}, \mathbf{z}} \left[ \| \mathbf{s}_\theta(\mathbf{x}_t, \mathbf{y}, t) + \frac{\mathbf{z}}{\sigma(t)} \|_2^2 \right],
\end{equation}
where $\mathbf{z} \sim \mathcal{N}(0, \mathbf{I})$ and $\mathbf{x}_t = \mu(\mathbf{x}_0, \mathbf{y}, t) + \sigma(t)\mathbf{z}$.

\subsection{Comparison: OUVE vs. BBED}

\paragraph{OUVE formulation:} The drift and diffusion coefficients for OUVE are defined as:
\begin{equation}
    \mathbf{f}(\mathbf{x}_t, t) = \gamma(\mathbf{y} - \mathbf{x}_t), \quad g(t) = \sqrt{ck^t},
\end{equation}
where $\gamma > 0$ is the {stiffness parameter} that controls the speed of the mean transition from $\mathbf{x}_0$ to $\mathbf{y}$. The parameters {$c > 0$ and $k > 1$ are hyperparameters governing the noise schedule}, where $c$ scales the diffusion intensity and $k$ determines the exponential growth rate of the noise variance over time. Under these definitions, the closed-form mean and variance are:
\begin{align}
    \mu(\mathbf{x}_0, \mathbf{y}, t) &= e^{-\gamma t}\mathbf{x}_0 + (1 - e^{-\gamma t})\mathbf{y}, \\
    \sigma^2(t) &= \frac{c(k^{t} - e^{-2\gamma t})}{2\gamma + \log k}.
\end{align}
% \textit{Note: Ensure the exponent in Eq. 15 matches your $g(t)$ definition. If $g(t) = \sqrt{ck^t}$, the variance typically involves $k^t$. If your implementation uses $k^{2t}$, then $g(t)$ should be $\sqrt{ck^{2t}}$.}

\paragraph{BBED formulation:} BBED resolves the prior mismatch by redefining the drift coefficient:
\begin{equation}
    \mathbf{f}(\mathbf{x}_t, t) = \frac{\mathbf{y} - \mathbf{x}_t}{1 - t},
\end{equation}
while maintaining the same diffusion coefficient $g(t)$ to ensure a consistent noise schedule for fair comparison. The linear mean evolution is given by:
\begin{equation}
    \mu(\mathbf{x}_0, \mathbf{y}, t) = (1 - t)\mathbf{x}_0 + t\mathbf{y}.
\end{equation}
As $t \to 1$, the mean exactly reaches $\mathbf{y}$, effectively eliminating the prior mismatch encountered in OUVE. To ensure numerical stability and avoid the singularity at $t=1$, we set the terminal diffusion time $T=0.999$.

\subsection{Sampling via Euler-Maruyama}
During inference, we discretize the interval $[0, T]$ into $N$ sub-intervals with step size $\Delta t = T/N$. The reverse process is approximated using the Euler-Maruyama method:
\begin{equation}
    \mathbf{x}_{t-\Delta t} = \mathbf{x}_t - [-\mathbf{f}(\mathbf{x}_t, t) + g^2(t)\mathbf{s}_\theta(\mathbf{x}_t, \mathbf{y}, t)]\Delta t + g(t)\sqrt{\Delta t} \mathbf{z},
\end{equation}
where $\mathbf{z} \sim \mathcal{N}(0, \mathbf{I})$. This approach allows us to generate clean samples starting from the initial state $\mathbf{x}_T \sim \mathcal{N}(\mathbf{y}, \sigma^2(T)\mathbf{I})$.

\begin{table}[htbp]
\centering
\caption{Distortion categories and corresponding probabilities in the synthetic training set.}
\label{tab:distortions}
\footnotesize % 保持较小字体
\setlength{\tabcolsep}{6pt} % 适当调整列间距
\renewcommand{\arraystretch}{0.85} % 进一步压缩行高，使表格不显长
\begin{tabular}{llc}
\toprule
\textbf{Family} & \textbf{Distortion Type} & \textbf{Prob.} \\ \midrule
\multirow{2}{*}{Noise} & General additive noise & 0.50 \\
 & Gaussian white noise & 0.60 \\ \midrule
Reverb & Freeverb / RIR convolution & 0.20 \\ \midrule
\multirow{2}{*}{Microphone} & Bandpass filtering & 0.20 \\
 & Bad mic frequency response & 0.50 \\ \midrule
\multirow{3}{*}{ADC/DAC} & High-pass filtering & 0.70 \\
 & Low-pass filtering & 0.70 \\
 & Bit depth reduction (8-24 bit) & 0.10 \\ \midrule
\multirow{2}{*}{AGC} & Dynamic range expansion & 1.00 \\
 & Post-processing gain & 0.10 \\ \midrule
\multirow{4}{*}{Preprocessing} & Hard clipping (Hard limit) & 0.25 \\
 & Post-clipping gain change & 0.25 \\
 & Resampling (3-32 kHz) & 0.40 \\
 & Multi-algorithm resampling & 1.00 \\ \midrule
Transmission & GSM network compression & 0.25 \\ \midrule
\multirow{4}{*}{Misc.} & Speaker gain fluctuation & 0.20 \\
 & Nearend gain change & 0.20 \\
 & Phaser (Phase distortion) & 0.02 \\
 & Non-linear Tanh distortion & 0.01 \\ \bottomrule
\end{tabular}
\end{table}

\section{Design Philosophy of the TLB Strategy}
\label{app:TLB_philosophy}

\subsection{Motivation for Frequency-Band Partitioning}

\begin{figure*}[h]
\begin{minipage}[b]{1.0 \linewidth}
 \centering
 % \hspace{3cm}  % 向you移动7厘米
 \centerline{\includegraphics[width= \columnwidth]{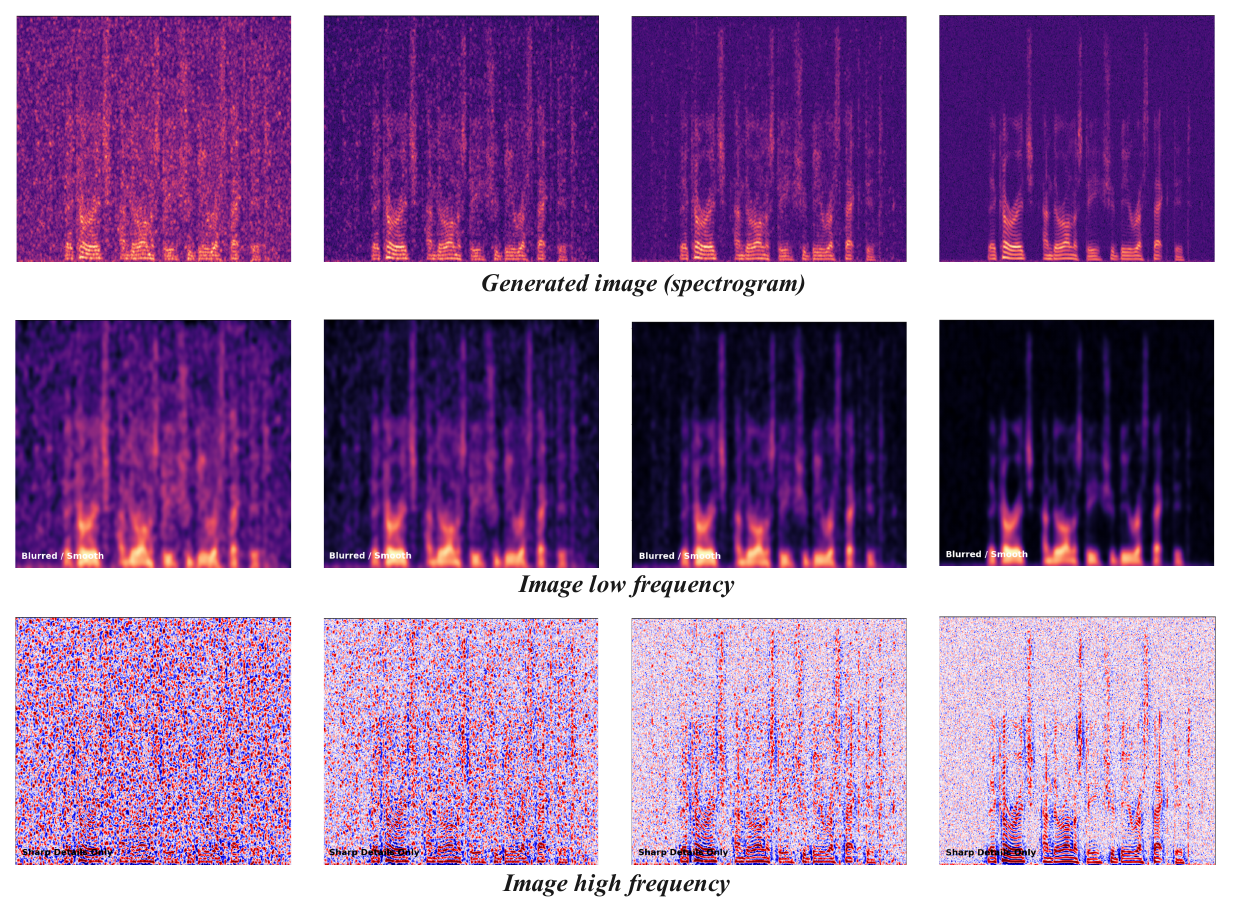}}
\end{minipage}

% \vspace{-0.5cm}
\caption{The reverse denoising process. The top row illustrates the image’s progressive denoising process across iterations, while the subsequent two rows display low-frequency and high-frequency components. }

\label{f:figstep_step}

\end{figure*}

The core logic for splitting the scaling parameters $s$ and $b$ into ``high'' and ``low'' frequency components is rooted in the distinct denoising dynamics of spectrograms when treated as 2D images. As illustrated in Fig.~\ref{f:figstep_step}, during the reverse diffusion process, high-frequency regions typically exhibit rapid denoising, whereas low-frequency structural components evolve more gradually. From an acoustic perspective, we define high-quality restoration via two dimensions:

\begin{itemize}
    \item \textbf{Harmonic Integrity (The ``Comb'' Structure):} In the low-to-mid frequency regions, clean speech is characterized by sharp, horizontal ``comb-like'' stripes representing the fundamental frequency ($f_0$) and its harmonics. Clear energy valleys (low-energy gaps) between these stripes are indicative of superior pitch restoration. 
    \item \textbf{Consonant Texture (High-Frequency Transients):} High-frequency regions (the upper portion of the Y axis) consist of vertical, stochastic yet bounded energy blocks representing consonants. Precise reconstruction requires sharp temporal boundaries and distinct textures rather than smeared background noise.
\end{itemize}

We empirically set {2\,kHz} \cite{schroter2022deepfilternet, yu2023efficient} (corresponding to {Frequency Bin 64} in our 512 points STFT) as the boundary. This threshold is chosen because the 0--2\,kHz range encompasses the $f_0$ and the majority of critical formants, accounting for over 80\% of the total spectral energy and speech information. By splitting the parameters at this boundary, TLB can independently recalibrate the backbone's high-frequency suppression and the skip-connections' low-frequency structural compensation.

\subsection{Parameter Selection and Branch-Specific Application}
Our 3-layer U-Net backbone consists of two encoder-decoder levels and a bottleneck. The TLB strategy is applied to the first two levels, resulting in a total of eight parameters: 
\begin{itemize}
    \item \textbf{Level 1:} $s_{1,\text{high}}, s_{1,\text{low}}$ (skip) and $b_{1,\text{high}}, b_{1,\text{low}}$ (backbone).
    \item \textbf{Level 2:} $s_{2,\text{high}}, s_{2,\text{low}}$ (skip) and $b_{2,\text{high}}, b_{2,\text{low}}$ (backbone).
\end{itemize}

Notably, the {TLB strategy is exclusively applied to the diffusion branch}. Although both the predictive and diffusion branches utilize U-Net architectures, the predictive branch processes complex-valued spectrograms (incorporating real, imaginary, and magnitude components). Direct numerical modulation of these feature maps during inference can inadvertently disrupt the delicate relationship between phase and magnitude, leading to reconstructive errors and performance degradation. In contrast, the diffusion branch operates on a score-based manifold that is better suited for such spectral-image recalibration, allowing TLB to enhance perceptual quality without compromising phase consistency.
Specifically, a value of $s$ \textbf{exceeding unity} ($s > 1$) amplifies the injection of high-frequency details from the skip-connections, whereas a value \textbf{below unity} ($s < 1$) attenates this contribution, reducing the fine-grained spectral textures. Similarly, $b$ regulates the backbone's denoising intensity: $b > 1$ {enhances} the noise suppression capability, while $b < 1$ {mitigates} the denoising effect to prevent over-smoothing. Therefore, the strategic adjustment of $s$ and $b$ is crucial for the TLB strategy to adaptively balance detail restoration and interference rejection across varying signal-to-noise ratios.

\begin{algorithm}[t]
\caption{Inference of DVPD with TLB Strategy}
\label{alg:dvpd_inference}
\begin{algorithmic}[1]
\STATE {\bfseries Input:} Degraded components $\{Y_r, Y_i, Y_m\}$; Starting time $T_{rs}$; Steps $N$; Fusion factor $\alpha$; TLB factors $\mathcal{S} = \{s_{l,band}\}$, $\mathcal{B} = \{b_{l,band}\}$.
\STATE {\bfseries Output:} Enhanced components $\{\hat{X}_r, \hat{X}_i\}$.

\STATE \textit{Step 1: Predictive Prior Generation}
\STATE $[\bar{X}^{p}_r, \bar{X}^{p}_i], \{h^{p}_l\}_{l=0}^5 \leftarrow P_\theta([Y_r, Y_i, Y_m])$ 
\STATE $\bar{X}^{p}_m \leftarrow \sqrt{(\bar{X}^{p}_r)^2 + (\bar{X}^{p}_i)^2}$
\STATE $\Phi^{p} \leftarrow \text{atan2}(\bar{X}^{p}_i, \bar{X}^{p}_r)$

\STATE \textit{Step 2: Diffusion Initialization}
\STATE $X_{T_{rs}} \leftarrow \mu(\bar{X}^{p}_m, Y_m, T_{rs}) + \sigma(T_{rs})\mathbf{Z}, \quad \mathbf{Z} \sim \mathcal{N}(0, \mathbf{I})$

\STATE \textit{Step 3: Generative Refinement with TLB}
\FOR{$t = T_{rs}, T_{rs}-\Delta t, \dots, \Delta t$}
    \STATE \textit{Inside $G_\theta(X_t, \{h^{pred}_l\}, t)$ with TLB recalibration:}
    \FOR{Level $l \in \{1, 2\}$}
        \STATE $x_{skip,l} \leftarrow \text{Encoder}_l(X_t)$
        \STATE $x_{skip,l}(\mathcal{F}_{low}) \leftarrow x_{skip,l}(\mathcal{F}_{low}) \cdot s_{l,low}$ 
        \STATE $x_{skip,l}(\mathcal{F}_{high}) \leftarrow x_{skip,l}(\mathcal{F}_{high}) \cdot s_{l,high}$
    \ENDFOR
    
    \FOR{Level $l \in \{1, 2\}$}
        \STATE $x_{back,l} \leftarrow \text{Decoder}_l(\cdot)$
        \STATE $\mathbf{m}_l \leftarrow \text{Normalize}(\text{Mean}(x_{back,l}))$ 
        \STATE $x_{back,l}(\mathcal{F}_{low}) \leftarrow x_{back,l}(\mathcal{F}_{low}) \odot ((b_{l,low}-1) \cdot \mathbf{m}_l + 1)$
        \STATE $x_{back,l}(\mathcal{F}_{high}) \leftarrow x_{back,l}(\mathcal{F}_{high}) \odot ((b_{l,high}-1) \cdot \mathbf{m}_l + 1)$
    \ENDFOR
    
    \STATE $s_\theta \leftarrow \text{Score}(x_{back,1}, \dots)$
    \STATE $X_{mean} \leftarrow X_t + [-f(X_t, t) + g(t)^2 s_\theta]\Delta t$
    \STATE $X_{t-\Delta t} \leftarrow X_{mean} + g(t)\sqrt{\Delta t}\mathbf{Z}$
\ENDFOR

\STATE \textit{Step 4: Synergistic Magnitude Fusion}
\STATE $\hat{X}^{d}_m \leftarrow \text{Clip}(X_{mean}, 0, +\infty)$
\STATE $\hat{X}_m \leftarrow \alpha \bar{X}^{p}_m + (1-\alpha)\hat{X}^{d}_m$

\STATE \textit{Step 5: Waveform Reconstruction}
\STATE $\hat{X}_r, \hat{X}_i \leftarrow \hat{X}_m \cos(\Phi^{p}), \hat{X}_m \sin(\Phi^{p})$
\STATE \textbf{return} $\{\hat{X}_r, \hat{X}_i\}$
\end{algorithmic}
\end{algorithm}

\section{Detailed Dataset Descriptions}
\label{app:datasets}

In this section, we provide the implementation details and synthesis protocols for the datasets used in our evaluation.

\textbf{WSJ0-UNI:}
Following the pipeline established by the Speech Signal Improvement Challenge \cite{ristea2025icassp}, we synthesized the WSJ0-UNI dataset to evaluate the model's capacity for universal restoration. We extended the diversity of distortions compared to previous works to include: recorded ambient noise, room reverberation, microphone frequency response variations, analog-to-digital converter (ADC) effects, automatic gain control (AGC) artifacts, and transmission-induced signal degradations (Table~\ref{tab:distortions}). Clean speech are sourced from the Wall Street Journal (WSJ0) corpus \cite{garofolo2007csr}. We utilized the \texttt{si\_tr\_s} set for training, and \texttt{si\_dt\_05} and \texttt{si\_et\_05} for validation and testing, respectively. Noise clips were randomly sampled from the WHAM! dataset \cite{wichern2019wham}.

\textbf{VoiceBank+DEMAND (VBDMD)}
As a standard benchmark for monaural speech denoising, we utilized the publicly available VBDMD dataset \cite{botinhao2016investigating}. The training set includes 11,572 utterances from 28 speakers, while the test set consists of 872 utterances from 2 distinct speakers (non-overlapping with the training set). Clean signals were mixed with DEMAND \cite{thiemann2013diverse} noise at Signal-to-Noise Ratios (SNRs) of $\{0, 5, 10, 15\}$\,dB for training and $\{2.5, 7.5, 12.5, 17.5\}$\,dB for testing. 

\textbf{VBDMD-REVERB (VBD-RB)}
To assess the model's generalization to unseen reverberant environments, we created the VBDMD-REVERB set by applying a stereo reverberation algorithm \cite{schroeder2003colorless} to the clean test utterances of VBDMD. The resulting average reverberation time ($T_{60}$) is 0.4\,s, providing a robust evaluate for dereverberation capabilities without specific training on this set.

\textbf{VBDMD-SR (Bandwidth Extension)}
Speech Super-Resolution (SR), or bandwidth extension, requires the model to reconstruct high-frequency spectral content from band-limited signals. We simulated this by applying a 12th-order Butterworth low-pass filter with a cutoff frequency of 4\,kHz to the VB-DMD test set. This task evaluates the generative strength of our dual-view framework in recovering high-frequency harmonic structures.

% \textbf{WSJ0-CHiME3 (WSJ0-CE3)}
% To verify the model's robustness against out-of-distribution real-world noise, clean WSJ0 utterances were combined with noise recordings from the CHiME3 dataset \cite{barker2015third}. The mixing SNRs were uniformly sampled between $-6$ and $14$\,dB, representing severe acoustic conditions encountered in urban environments.

\paragraph{WSJ0-CHiME3 (WSJ0-CE3)} 
To rigorously assess the model's robustness against {out-of-distribution (OOD)} real-world noise, we synthesized the WSJ0-CE3 benchmark by combining clean utterances with authentic urban noise recordings.
The clean components are sourced from the Wall Street Journal (WSJ0) corpus \cite{garofolo2007csr}, a gold-standard benchmark in speech processing. It consists of high-fidelity, read-out news text recorded in controlled, anechoic-like environments using close-talk microphones. This ensures that the base signal is devoid of any pre-existing background noise or reverberation, providing a clear ground truth for reconstruction.
The noise signals are extracted from the third CHiME challenge (CHiME-3) dataset \cite{barker2015third}. Unlike synthetic or stationary noise, CHiME-3 features multi-channel recordings captured in four highly dynamic and challenging urban environments: \textit{public transport (Bus)}, \textit{busy cafes (Cafe)}, \textit{pedestrian areas (Pedestrian)}, and \textit{street junctions (Street)}. These recordings contain highly non-stationary acoustic events, such as background chatter, engine rumbles, and sudden clatter, which are representative of the complex distortions encountered in daily communication.
For our evaluation, we randomly selected clean utterances from the WSJ0 \texttt{si\_et\_05} set and mixed them with noise clips from the CHiME-3 test partition. To simulate varying levels of signal degradation, the Signal-to-Noise Ratio (SNR) was uniformly sampled from a wide range of {$-6$ to $14$~dB}. This setup, particularly at negative SNR levels, represents severe acoustic conditions that demand high generative capability to restore intelligibility. All mixed signals were resampled to 16\,kHz.

\textbf{WSJ0-REVERB (WSJ0-RB)}
We simulated realistic reverberant scenarios using the \textit{pyroomacoustics} engine. Room dimensions were uniformly sampled within $[5, 15] \times [5, 15] \times [2, 6]$\,m, with $T_{60}$ values ranging from 0.4 to 1.0\,s. This resulted in a mean Direct-to-Reverberant Ratio (DRR) of approximately $-9$\,dB and a measured average $T_{60}$ of 0.91\,s.

\section{Training Objectives and Evaluation Metrics}
\label{app:loss_and_metrics}

\subsection{Loss Functions for the Predictive Branch}
To ensure the predictive branch provides high-quality phase and magnitude estimates, we employ a combination of spectral and phase-aware losses.

\paragraph{Spectral Losses} The magnitude and complex-domain Mean Square Errors (MSE) are defined as:
\begin{align}
    \mathcal{L}_{\text{mag}} &= \mathbb{E} [\| \hat{X}_{\text{pred},m} - X_m \|_2^2], \\
    \mathcal{L}_{\text{comp}} &= \mathbb{E} [\| \hat{X}_{\text{pred},r} - X_r \|_2^2 + \| \hat{X}_{\text{pred},i} - X_i \|_2^2].
\end{align}

\paragraph{Anti-wrapping Phase Loss} Given that the final enhanced waveform inherits the phase from the predictive branch, we introduce an anti-wrapping phase loss $\mathcal{L}_{\text{pha}}$ \cite{lay2023reducing} to mitigate phase discontinuities:
\begin{equation}
    \mathcal{L}_{\text{pha}} = \mathcal{L}_{\text{IP}} + \mathcal{L}_{\text{GD}} + \mathcal{L}_{\text{IAF}},
\end{equation}
where $\mathcal{L}_{\text{IP}}$, $\mathcal{L}_{\text{GD}}$, and $\mathcal{L}_{\text{IAF}}$ denote the instantaneous phase, group delay, and instantaneous angular frequency losses. To handle the periodicity of phase, we utilize the anti-wrapping function $f_{\text{AW}}(t) = |t - 2\pi \cdot \text{round}(t/2\pi)|$. These losses are computed using differential operators along the frequency and time axes to ensure spectro-temporal phase consistency.

\subsection{Detailed Definitions of Evaluation Metrics}
\label{app:metrics_detail}

To provide a multifaceted evaluation of the proposed DVPD, we employ a suite of objective metrics encompassing perceptual quality, intelligibility, signal fidelity, and deep learning based subjective score estimation.

\paragraph{PESQ (Perceptual Evaluation of Speech Quality):} 
Validated by ITU-T Recommendation P.862 \cite{rix2001perceptual}, PESQ is the most widely used intrusive metric for assessing the quality of narrow-band and wide-band speech. It models the human auditory system by comparing the loudness spectra of the reference and degraded signals. In this paper, we report the Wideband PESQ (P.862.2) scores, ranging from {0.5 to 4.5}, where higher scores indicate superior perceptual quality.

\paragraph{ESTOI (Extended Short-Time Objective Intelligibility):} 
ESTOI \cite{jensen2016algorithm} is an extension of the STOI metric designed to predict the intelligibility of speech in the presence of highly non-stationary noise. Unlike STOI, ESTOI accounts for the dependencies between frequency bands, making it more robust for evaluating complex restoration tasks. The score ranges from {0 to 1}, representing the percentage of words likely understood by a human listener.

\paragraph{SI-SDR (Scale-Invariant Signal-to-Distortion Ratio):} 
SI-SDR \cite{le2019sdr} is an energy-ratio-based metric that measures global reconstruction fidelity. Crucially, it is invariant to the gain difference between the estimated and reference signals, which is particularly important for generative models that may produce variations in signal amplitude. It is defined as:
\begin{equation}
    \text{SI-SDR} = 10 \log_{10} \left( \frac{\| \mathbf{s}_{\text{target}} \|^2}{\| \mathbf{e}_{\text{noise}} \|^2} \right),
\end{equation}
where $\mathbf{s}_{\text{target}}$ is the orthogonal projection of the estimated signal onto the clean reference.

\paragraph{WV-MOS (Wav2vec MOS):} 
WV-MOS \cite{andreev2023hifi++} is a non-intrusive metric based on a pre-trained {Wav2Vec 2.0} model. It is trained on large-scale subjective datasets to predict human Mean Opinion Scores directly from raw waveforms. It has shown a high correlation with subjective listening tests in recent generative speech enhancement literature.

\paragraph{DNS-MOS (Deep Noise Supression MOS):} 
The DNS-MOS metric \cite{reddy2021dnsmos} was developed by Microsoft based on the ITU-T P.808 standard. It utilizes a deep neural network to predict subjective quality for noise suppression tasks. DNS-MOS provides a robust estimation of perceptual quality in diverse real-world environments, making it a critical benchmark for the Universal Speech Enhancement (USE) scenarios investigated in this work.

\paragraph{Composite Metrics (CSIG, CBAK, COVL):} 
Following the methodology in \cite{hu2007evaluation}, we report three composite Mean Opinion Score (MOS) estimates:
\begin{itemize}
    \item \textbf{CSIG:} Predicts the MOS of signal distortion, focusing on the integrity of the speech itself.
    \item \textbf{CBAK:} Predicts the MOS of background noise intrusiveness, measuring the suppression of residual noise.
    \item \textbf{COVL:} Predicts the overall MOS, representing the comprehensive quality of the enhanced audio.
\end{itemize}
All composite metrics range from {1 to 5}.

\section{Sensitivity Analysis of Inference Hyperparameters}
\label{app:inference_ablation}

In this section, we provide detailed ablation studies on the key inference hyperparameters: the weighting factor $\alpha$, the starting time step $T_{rs}$, and the number of reverse sampling steps $N$. All experiments in this section were conducted on the WSJ0-UNI test set.

\subsection{Balance between Predictive and Generative Branches ($\alpha$)}
The parameter $\alpha$ weights the magnitude spectra from the predictive ($\hat{\mathbf{X}}_{\text{p},m}$) and diffusion ($\hat{\mathbf{X}}_{\text{d},m}$) branches. As shown in Table \ref{tab:alpha_ablation}, when $\alpha=1.0$, the model relies solely on the predictive branch, leading to higher SI-SDR but lower PESQ due to spectral over-smoothing. Conversely, a very small $\alpha$ (e.g., $0.2, 0$) introduces more generative details but slightly degrades signal fidelity. We empirically set $\alpha=0.4$ as it yields the optimal balance between perceptual quality (PESQ) and objective fidelity.

\begin{table}[h]
\centering
\caption{Ablation of the weighting factor $\alpha$ ($T_{rs}=0.12, N=3$).}
\label{tab:alpha_ablation}
\begin{tabular}{lccccc}
\toprule
$\alpha$ & PESQ $\uparrow$ & SI-SDR $\uparrow$ & ESTOI $\uparrow$ & WV-MOS $\uparrow$ & DNS-MOS $\uparrow$ \\ \midrule
1.0 (Pred. Only) & 2.70 & \textbf{20.55} & 0.88 & 3.76 & 3.42 \\
0.6 & 2.88 & 20.03 & 0.88 & 3.91 & 3.48 \\
\textbf{0.4 (Ours)} & \textbf{2.99} & 19.15 & \textbf{0.91} & {4.16} & \textbf{3.51} \\
0.2 & 2.78 & 19.98 & 0.88 & 4.18 & 3.50 \\
0.0(Diff. Only) & 2.71 & 18.77 & 0.89 & \textbf{4.23} & 3.48 \\\bottomrule
\end{tabular}
\end{table}

\subsection{Impact of Sampling Steps ($N$)}
We first evaluate the impact of the number of reverse steps $N$ by fixing $T_{rs} = T$ and $\alpha=0.4$. As shown in Table \ref{tab:n_ablation}, both perceptual metrics (PESQ, DNS-MOS) and intelligibility (ESTOI) improve significantly as $N$ increases from 10 to 25. However, beyond $N=25$, the performance gains plateau and even exhibit a slight regression in WV-MOS. This suggests that 25 steps are sufficient to capture the generative distribution of clean speech, and further iterations only increase computational MACs without providing meaningful acoustic refinement.

\begin{table}[h]
\centering
\caption{Ablation of sampling steps $N$ (fixing $\alpha=0.4, T_{rs}=T$).}
\label{tab:n_ablation}
\resizebox{0.7\columnwidth}{!}{%
\begin{tabular}{lccccc}
\toprule
$N$ & PESQ $\uparrow$ & SI-SDR $\uparrow$ & ESTOI $\uparrow$ & WV-MOS $\uparrow$ & DNS-MOS $\uparrow$ \\ \midrule
10  & 2.88 & \textbf{20.45} & 0.89 & 3.73 & 3.42 \\
15  & 2.93 & 21.20 & 0.90 & 3.98 & 3.51 \\
20  & 2.98 & 21.12 & 0.90 & 4.19 & 3.55 \\
\textbf{25} & \textbf{3.08} & 20.11 & \textbf{0.92} & {4.33} &{3.58} \\
30  & 3.11 & 19.01 & \textbf{0.92} & 4.34 &  \textbf{3.60} \\
35  & 3.09 & 19.28 & \textbf{0.92} & 4.32 &  \textbf{3.60} \\
40  & 3.11 & 18.99 & \textbf{0.92} & \textbf{4.37} & 3.58 \\ \bottomrule
\end{tabular}%
}
\end{table}

\subsection{Impact of Trajectory Starting Point ($T_{rs}$)}

We further investigate the effect of the starting time $T_{rs}$ with a fixed discretization step $\Delta t = 0.04$ and $\alpha = 0.4$. As illustrated in Table \ref{tab:trs_ablation}, PESQ continues to show marginal improvements as $T_{rs}$ increases up to 0.24, indicating that a longer diffusion path helps in reconstructing fine-grained spectral details. However, WV-MOS and DNS-MOS reach their peak at $T_{rs}=0.12$. When $T_{rs} > 0.12$, we observe a gradual decline in these metrics, as the stochastic nature of the diffusion process begins to deviate from the ground-truth waveform. Thus, $T_{rs}=0.12$ is selected as the optimal starting point to balance generative realism and signal fidelity.

\begin{table}[h]
\centering
\resizebox{0.7\linewidth}{!}{%
\begin{tabular}{lccccc}
\toprule
$T_{rs}$ & PESQ $\uparrow$ & SI-SDR $\uparrow$ & ESTOI $\uparrow$ & WV-MOS $\uparrow$ & DNS-MOS $\uparrow$ \\ \midrule
0.04 & 2.61 & \textbf{19.45} & 0.87 & 3.76 & 3.36 \\
0.08 & 2.78 & 19.40 & 0.89 & 3.98 & 3.46 \\
\textbf{0.12} & \textbf{2.99} & 19.15 & {0.91} & \textbf{4.16} & \textbf{3.51} \\
0.16 & 3.06 & 18.89 & 0.91 & 4.11 & 3.48 \\
0.20 & 3.10 & 18.71 & \textbf{0.92} & 4.07 & 3.48 \\
0.24 & \textbf{3.13} & 18.62 & \textbf{0.92} & 3.99 & 3.44 \\ \bottomrule
\end{tabular}%
}
\caption{Ablation of starting time $T_{rs}$ (fixing $\alpha=0.4, \Delta t=0.04$).}
\label{tab:trs_ablation}
\end{table}

\section{Detailed Configuration of Ablation Variants}
\label{app:ablation_config}

\begin{figure*}[t]
    \centering 
    \includegraphics[width=0.85\textwidth]{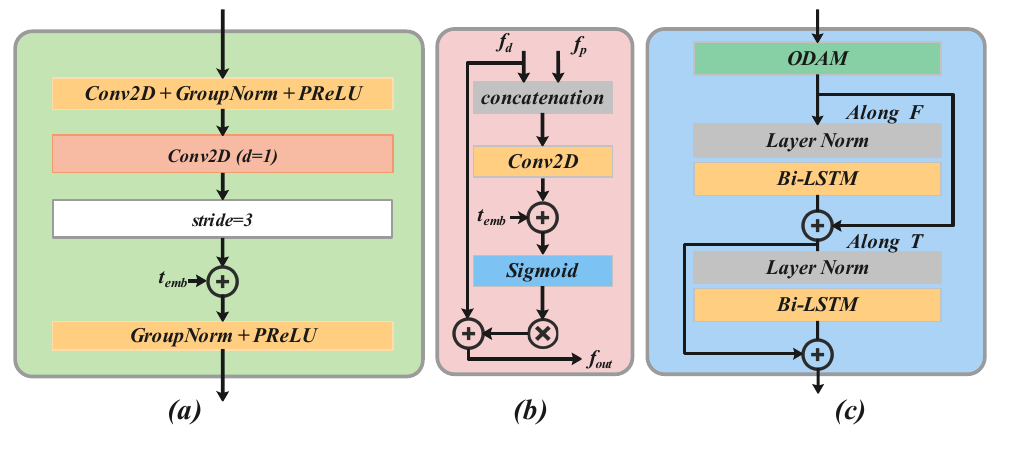}
    \caption{{Architectural modifications for the ablation study.} (a) The simplified encoder replacing {FANC}, using a uniform stride without band splitting. (b) The {FI} variant with the frequency-selective branch disabled. (c) The baseline module replacing {LISA}, featuring a standard dual-path processing structure.}
    \label{f:ablation_module}
\end{figure*}

To further clarify the contribution of each module, we provide the implementation details for the ablation variants discussed in Section~\ref{sec:ablation_study}. The architectural modifications are illustrated in Fig.~\ref{f:ablation_module}.

\begin{itemize}
    \item \textbf{w/o FANC Encoder:} In this variant, the non-uniform spectral partitioning strategy is disabled. Instead of processing the three frequency bands independently with heterogeneous dilated kernels, we employ a standard 2D convolution with a uniform stride of 3 across the entire frequency axis, as shown in Fig.~\ref{f:ablation_module}(a).
    
    \item \textbf{w/o FI Module:} The Frequency-Aware Interaction (FI) module is not entirely removed; rather, we deactivate the frequency-selective subpath. In this configuration, the predictive branch provides guidance to the diffusion branch without adaptive band-wise weighting, as depicted in Fig.~\ref{f:ablation_module}(b).
    
    \item \textbf{w/o Phase Loss:} For this experiment, the anti-wrapping phase regularization term $\mathcal{L}_{\text{pha}}$ is excluded from the optimization objective. The resulting total loss is formulated as:
    \begin{equation}
        \mathcal{L} = \lambda_1 \mathcal{L}_{\text{mag}} + (1 - \lambda_1) \mathcal{L}_{\text{comp}} + \mathcal{L}_{\text{score}}.
    \end{equation}
    
    \item \textbf{w/o LISA Module:} The multi-range parallel branches and dynamic filtering mechanism are removed. We replace the LISA module with a conventional dual-path architecture (e.g., intra- and inter-block processing) commonly used in standard speech models \cite{luo2020dual, zhang2025composite}, as illustrated in Fig.~\ref{f:ablation_module}(c).
\end{itemize}

\section{Performance Analysis of TLB across Quality-based Stratifications}
\label{app:Performance Analysis of TLB}

In this section, we provide a detailed rationale for stratifying the test samples based on their baseline PESQ scores and present the experimental validation for our parameter selection. In a standard U-Net architecture, the scaling parameters $(s, b)$ are expected to maintain a delicate equilibrium: $s$ is designed to compensate for spectral energy (particularly in high-frequency details), while $b$ aims to suppress residual noise and preserve low-frequency speech integrity.

\subsection{Case I: Severely Degraded Samples ($\text{PESQ} < 2$)}
\label{subsec:tier1}

\begin{figure*}[t]
    \centering 
    \includegraphics[width=\textwidth]{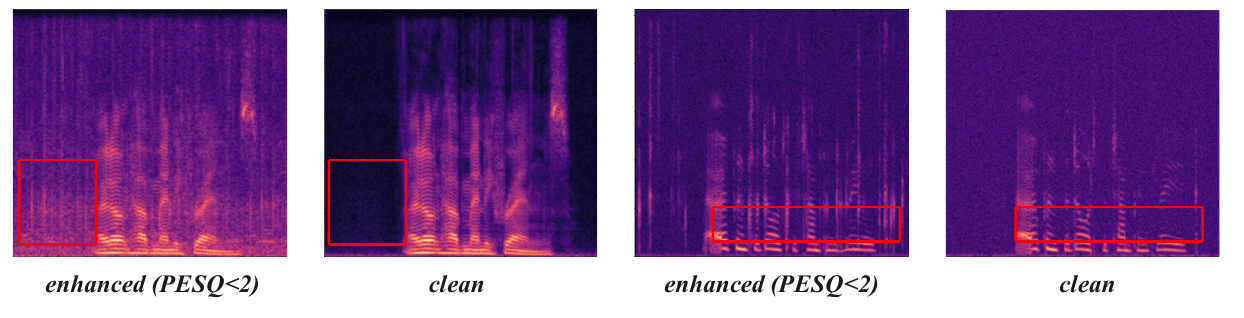}
    \caption{Visualization of two typical artifacts in enhanced speech with $\text{PESQ} < 2$. \textbf{Left}: High-intensity residual noise localized in low-frequency quasi-silent regions. \textbf{Right}: Deficiency of core low-frequency bands caused by the over-suppression of speech energy.}
    \label{f:pesq2}
\end{figure*}

For samples in the lowest quality tier ($\text{PESQ} < 2$), the primary challenge lies in rescuing speech information from severe corruption. As illustrated in Fig.~\ref{f:pesq2}, the restored spectrograms from the baseline model typically exhibit two critical artifacts:
\begin{enumerate}
    \item \textbf{Over-suppression in Low-frequency Bands:} Since low-frequency components contain the majority of speech energy and phonetic information, the model's tendency to over-denoise leads to a significant loss of intelligibility.
    \item \textbf{Residual Noise in Quasi-silent Regions:} Significant noise often persists in low-frequency intervals that should ideally be silent, which heavily penalizes the PESQ score.
\end{enumerate}

\paragraph{Parameter Configuration Logic:} 
To address these issues, our TLB strategy prioritizes the restoration of low-frequency spectral integrity. We observe that:
\begin{itemize}
    \item \textbf{Low-frequency Enhancement ($s_{1,low}, s_{2,low} \uparrow$):} We set these scaling factors to relatively large values. This provides the U-Net with additional high-frequency information and spectral energy in the low-frequency regions to counteract over-suppression.
    \item \textbf{Adaptive Noise Suppression ($b_{1,low}, b_{2,low} \uparrow$):} While increasing $s$ restores energy, it inevitably introduces extra noise artifacts. Therefore, the corresponding bias parameters must also be increased to suppress both the inherent residual noise and the artifacts introduced by $s$. 
    \item \textbf{The $b < s$ Constraint:} Crucially, we maintain the constraint that the bias values ($b_{1,low}, b_{2,low}$) do not exceed the scaling factors ($s_{1,low}, s_{2,low}$). This ensures that the denoising intensity does not outweigh the energy compensation, resulting in a net gain in speech quality.
\end{itemize}

In this case, since the overall quality is predominantly dictated by low-frequency restoration, the high-frequency parameters ($s_{high}, b_{high}$) can be assigned lower priority or reduced values to avoid introducing unnecessary high-frequency hiss. To validate this strategy, we conducted an extensive ablation study to identify the optimal balance for severely degraded samples. A representative optimal configuration identified through our grid search is:
\begin{itemize}
    \item \textbf{Low-frequency components:} $s_{1,low}=2.0, s_{2,low}=4.1$ and $b_{1,low}=2.5, b_{2,low}=1.5$.
    \item \textbf{High-frequency components:} $s_{1,high}=0.8, s_{2,high}=0.5$ and $b_{1,high}=1.0, b_{2,high}=1.0$.
\end{itemize}
The performance variations under different parameter combinations are illustrated in Table \ref{tab:ablation_low_freq}. These results confirm that aggressive scaling in the low-frequency bands, coupled with synchronized bias suppression, yields the most significant PESQ improvements for low-quality speech.

\begin{table}[htbp]
\centering
\caption{Ablation study of the four critical low-frequency parameters for Case 1 ($\text{PESQ} < 2$).}
\label{tab:ablation_low_freq}
\begin{tabular}{lcccc|cc}
\toprule
\textbf{Configuration} & $s_{1,low}$ & $b_{1,low}$ & $s_{2,low}$ & $b_{2,low}$ & \textbf{PESQ} & \textbf{STOI} \\ \midrule
Baseline (Original)    & 1.0 & 1.0 & 1.0 & 1.0 & 1.73 & 0.77 \\ \midrule
Only Level 1           & 1.5 & 2.0 & 1.0 & 1.0 & 1.75 & 0.77 \\
Only Level 1           & 1.8 & 2.3 & 1.0 & 1.0 & 1.75 & 0.77 \\
Only Level 1           & \textbf{2.0} & \textbf{2.5} & 1.0 & 1.0 & 1.76 & 0.77 \\
Only Level 1           & 2.2 & 2.7 & 1.0 & 1.0 & 1.75 & 0.77 \\

Only Level 2           & 1.0 & 1.0 & 3.5 & 1.3 & 1.78 & 0.77 \\
Only Level 2           & 1.0 & 1.0 & \textbf{4.1} & \textbf{1.5} & 1.81 & 0.77 \\
Only Level 2           & 1.0 & 1.0 & 4.5 & 1.7 & 1.73 & 0.77 \\\midrule
Only ($s$)             & 2.0 & 1.0 & 4.1 & 1.0 & 1.61 & 0.76 \\
Only ($b$)             & 1.0 & 2.5 & 1.0 & 1.5 & 1.71 & 0.77 \\ \midrule
\textbf{Optimal TLB}   & \textbf{2.0} & \textbf{2.5} & \textbf{4.1} & \textbf{1.5} & \textbf{1.88} & \textbf{0.78} \\ \bottomrule
\end{tabular}
\end{table}

\subsection{Case II: Moderately Restored Samples ($2 < \text{PESQ} \le 3$)}
\label{subsec:tier2}

\begin{figure*}[t]
    \centering 
    \includegraphics[width=\textwidth]{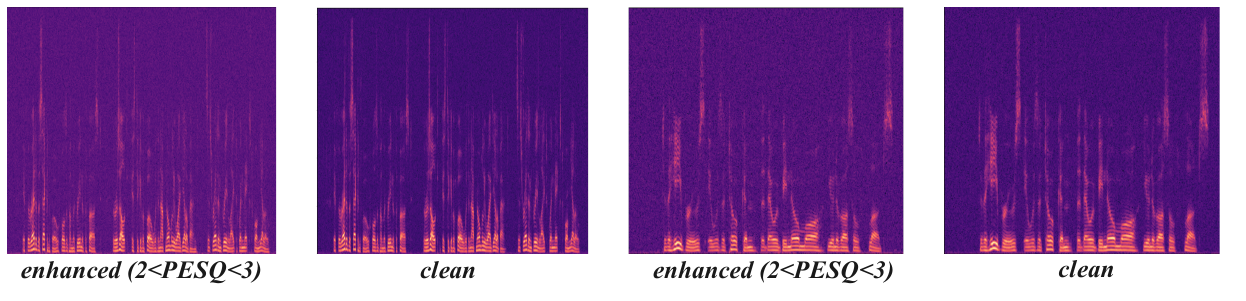}
    \caption{Visualization of two typical artifacts in enhanced speech with $2 < \text{PESQ} \le 3$.}
    \label{f:pesq2_3}
\end{figure*}

In the second tier, the restored speech generally exhibits high fidelity in the low-frequency regions with minimal structural degradation, contributing to a relatively high baseline PESQ. However, the primary remaining issue is the presence of subtle, quasi-uniform residual noise distributed across the entire spectro-temporal domain.

\paragraph{Parameter Configuration Logic:} 
Given that the artifacts are no longer concentrated in specific frequency bands, our strategy shifts toward global consistency. The parameter configuration follows these principles:
\begin{itemize}
    \item \textbf{Uniform Parameter Application:} To address the uniform distribution of noise, we synchronize the parameters across both high and low frequency bands, such that $s_{high} \approx s_{low}$ and $b_{high} \approx b_{low}$. This ensures a balanced restoration without introducing spectral tilts.
    \item \textbf{Aggressive Noise Suppression ($b \uparrow$):} The bias parameters are set to relatively large values to enhance the model's ability to suppress the persistent background hiss and "clear up" the global spectrogram.
    \item \textbf{Energy Compensation ($s > 1$):} Concurrently, the scaling factors ($s$) are also set greater than 1. This acts as a protective mechanism to prevent the aggressive bias term ($b$) from over-purifying the signal, which could otherwise lead to the loss of subtle speech components or "thinning" of the audio.
    \item \textbf{The $s < b$ Balance:} To ensure that the strategy remains effective for denoising, we maintain the constraint $s < b$. In this specific tier, this relationship ensures that the noise reduction intensity slightly outweighs the energy amplification, effectively lowering the noise floor without compromising the already well-restored speech structure.
\end{itemize}
In light of these principles, we adopt a frequency-agnostic configuration for this tier, applying identical parameters to both low and high-frequency bands to maintain spectral balance. Specifically, the optimized parameters are set as follows:
\begin{itemize}
    \item \textbf{Level 1:} $s_{1,low} = s_{1,high} = 1.5$ and $b_{1,low} = b_{1,high} = 2.5$.
    \item \textbf{Level 2:} $s_{2,low} = s_{2,high} = 2.0$ and $b_{2,low} = b_{2,high} = 1.5$.
\end{itemize}
This configuration reflects our strategy of aggressive noise suppression in the first stage ($b_1 > s_1$), followed by controlled energy recovery in the second stage ($s_2 > b_2$). To validate the efficacy of this cross-band uniform application and the selection of these specific values, we conducted a targeted ablation study, as summarized in Table \ref{tab:ablation_tier2}.

\begin{table}[htbp]
\centering
\caption{Ablation study of global uniform parameters for Case 2 ($2 < \text{PESQ} \le 3$). All parameters are synchronized across low and high frequencies ($s_{low}=s_{high}, b_{low}=b_{high}$).}
\label{tab:ablation_tier2}
\begin{tabular}{lcccc|cc}
\toprule
\textbf{Configuration} & $s_{1}$ & $b_{1}$ & $s_{2}$ & $b_{2}$ & \textbf{PESQ} & \textbf{STOI} \\ \midrule
Baseline (Original)    & 1.0 & 1.0 & 1.0 & 1.0 & 2.59 & 0.82 \\ \midrule
Level 1 Only           & 1.5 & 2.5 & 1.0 & 1.0 & 2.61 & 0.82 \\
Level 2 Only           & 1.0 & 1.0 & 2.0 & 1.5 & 2.62 & 0.82 \\ \midrule
\textbf{Uniform TLB (Optimal)} & \textbf{1.5*} & \textbf{2.5*} & \textbf{2.0*} & \textbf{1.5*} & \textbf{2.65} & \textbf{0.82} \\ \bottomrule
\end{tabular}
\begin{flushleft}
\footnotesize{*The optimal parameters ($s_1, b_1, s_2, b_2$) were identified through an exhaustive grid search and comprehensive empirical validation to ensure robust performance.}
\end{flushleft}
\end{table}

\subsection{Case III: High-Fidelity Samples ($\text{PESQ} \ge 3$)}
\label{subsec:tier3}

For samples with $\text{PESQ} \ge 3$, the restored speech already exhibits high fidelity, characterized by nearly intact spectral structures and negligible residual noise. In this high-quality regime, the focus of restoration shifts from fundamental reconstruction to the refinement of fine-grained spectral details.

\paragraph{Dominance of High-Frequency Components:}
We posit that for high-fidelity speech, the high-frequency regions play a decisive role in further improving perceptual quality. This is because:
\begin{itemize}
    \item \textbf{Low-frequency Saturation:} At this quality level, the low-frequency backbone of the speech is typically recovered to a near-optimal state. Continued intervention in the low-frequency bands yields diminishing returns and poses a risk of introducing unnecessary artifacts into an already stable signal.
    \item \textbf{Perceptual Brilliance:} The subjective clarity and naturalness of speech are highly sensitive to high-frequency harmonics and overtones. Minor spectral dampening or fine-grained noise in the high-frequency range becomes the primary bottleneck preventing the score from reaching a near-perfect level (e.g., PESQ > 4).
\end{itemize}

\paragraph{Parameter Configuration Logic:}
Based on the above rationale, we implement a selective optimization strategy:
\begin{itemize}
    \item \textbf{High-frequency Specific Tuning:} We exclusively apply the TLB strategy to the high-frequency scaling and bias parameters ($s_{high}, b_{high}$). By fine-tuning these values, we can subtly enhance the high-frequency energy and sharpen the spectral peaks, thereby improving the overall "crispness" of the audio.
    \item \textbf{Low-frequency Preservation:} The low-frequency parameters ($s_{low}, b_{low}$) are maintained at their identity values (i.e., $s=1, b=0$). This "hands-off" approach ensures that the high-quality low-frequency components, which are already well-restored by the U-Net, remain untouched, preserving the fundamental integrity of the speech.
\end{itemize}
Given that the baseline quality in this tier is already near-optimal, the numerical improvement afforded by the TLB strategy is relatively marginal (approximately 0.02 PESQ). Consequently, visual inspection of the spectrograms yields no discernible differences; thus, we omit the visual comparison for the sake of brevity. However, to achieve this subtle refinement in perceptual clarity, we identified the following specific parameter configuration:
\begin{itemize}
    \item \textbf{Low-frequency components (Preservation):} $s_{1,low} = 1.0, s_{2,low} = 1.0$ and $b_{1,low} = 1.0, b_{2,low} = 1.0$.
    \item \textbf{High-frequency components (Refinement):} $s_{1,high} = 1.1, s_{2,high} = 1.2$ and $b_{1,high} = 1.2, b_{2,high} = 1.1$.
\end{itemize}
In this configuration, the low-frequency parameters are kept at their identity mapping to safeguard the fundamental speech structure, while the high-frequency parameters are slightly elevated to enhance spectral brilliance. This selective boosting confirms that for high-fidelity audio, "less is more," and targeted high-frequency adjustment is sufficient to capture the remaining perceptual gains.

\section{Visual results}
\label{Visual results}

\begin{figure*}[t]
    \centering 
    \includegraphics[width=\textwidth]{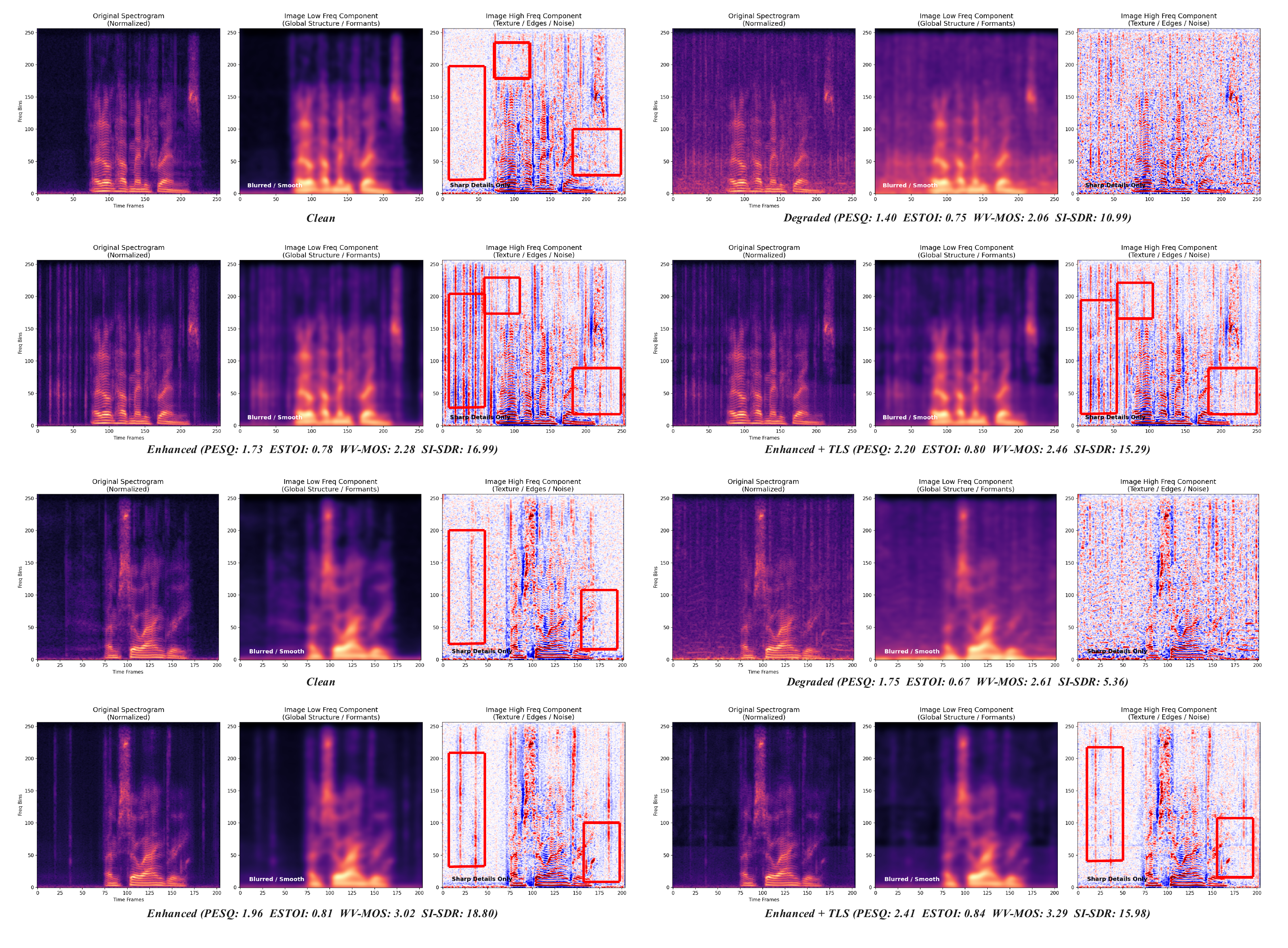}
    \caption{Visual comparison of spectrograms for clean speech, baseline enhanced speech, and speech refined by the TLB strategy. The red boxes highlight regions where the baseline model fails to suppress high-intensity noise streaks, whereas our TLB strategy effectively attenuates or eliminates these artifacts.}
\label{fig:spectrogram_comparison}
\end{figure*}

To provide a more intuitive demonstration of the effectiveness of the TLB strategy, we provide a comparative visualization of the spectrograms in Fig.~\ref{fig:spectrogram_comparison}. The red-boxed regions highlight the distinct differences between the clean reference, the baseline enhanced output, and the TLB result. 

It is observable that under conditions of severe degradation, the baseline enhancement model often fails to completely eliminate residual noise, which manifests as prominent \textit{vertical spectral streaks} (as shown in the red boxes). These artifacts represent structured noise that heavily degrades the perceptual quality and intelligibility. Upon applying the TLB strategy, these vertical structures are significantly attenuated or, in some cases, entirely removed. This qualitative refinement confirms that TLB not only improves numerical metrics but also effectively ``cleans" the spectro-temporal regions that are most challenging for standard U-Net architectures, leading to a substantial gain in overall speech quality.

\end{document}